\documentclass[a4paper,11pt]{article}
\usepackage{jheppub}
\usepackage[T1]{fontenc} % if needed
\usepackage{amssymb}
\usepackage{amsmath}
\usepackage{color}
\usepackage{xspace}
\usepackage{caption}
\usepackage{appendix}
\usepackage{float}
\usepackage{hyperref}
\usepackage{diagbox}
\usepackage{soul}
\usepackage{subfig}
\usepackage{slashed}
\usepackage{amsfonts}
\usepackage{multirow}
\usepackage{mathrsfs}
\usepackage{graphicx}
\usepackage{amsmath}
\usepackage{amssymb}
\usepackage{bm}
\usepackage{bbm}
\usepackage{color}
\usepackage{booktabs}

\newcommand{\nn}{\nonumber}
\newcommand{\bea}{\begin{align}}
	\newcommand{\eea}{\end{align}}
\newcommand{\beq}{\begin{equation}}
	\newcommand{\eeq}{\end{equation}}
\newcommand{\bqa}{\begin{eqnarray}}
	\newcommand{\eqa}{\end{eqnarray}}

\newcommand{\M}{\ensuremath \text{M}}

\newcommand{\eps}{\epsilon}

\newcommand{\als}{\alpha_s}
\newcommand{\bff}[1]{{\bf {#1}}}
\allowdisplaybreaks[4]

\title{ Analytic decay width of  the Higgs boson to massive bottom quarks at next-to-next-to-leading order in QCD}

\author[b,c]{Jian Wang,}
\author[a,b]{Yefan Wang\footnote{Corresponding author.},}
\author[b]{Da-Jiang Zhang}

\affiliation[a]{Department of Physics and Institute of Theoretical Physics, Nanjing Normal University,
Nanjing, Jiangsu 210023, China}
\affiliation[b]{School of Physics, Shandong University, Jinan, Shandong 250100, China}
\affiliation[c]{Center for High Energy Physics, Peking University, Beijing 100871, China}

\emailAdd{j.wang@sdu.edu.cn}
\emailAdd{wangyefan@sdu.edu.cn}
\emailAdd{zhangdajiang@mail.sdu.edu.cn}

\abstract{
The Higgs boson decay to a massive bottom quark pair provides the dominant contribution to the Higgs boson width.
We present an exact result for such a decay induced by the bottom quark Yukawa coupling with next-to-next-to-leading order (NNLO) QCD corrections.
We have adopted the canonical differential equations in the calculation and obtained the result in terms of multiple polylogarithms.
We also compute the contribution from the decay to four bottom quarks which consists of complete elliptic integrals or their one-fold integrals.
The result in the small bottom quark mass limit coincides with the previous calculation using the large momentum expansion.
The threshold expansion exhibits power divergent terms in the bottom quark velocity, which has a structure different from that in $e^+e^-\to t\bar{t}$ but can be reproduced by computing the corresponding Coulomb Green function.
The NNLO corrections significantly reduce the uncertainties from both the renormalization scale and the renormalization scheme of the bottom quark Yukawa coupling.
Our result can be applied to a heavy scalar decay to a top quark pair.

}

\begin{document}

\maketitle

\flushbottom

\section{Introduction}

The Higgs boson was discovered more than ten years ago, and its couplings with other elementary particles in the standard model (SM) have been measured and found to be consistent with the SM predictions \cite{ATLAS:2022vkf,CMS:2022dwd,ParticleDataGroup:2022pth}.
The Yukawa couplings of the Higgs boson, along with the spontaneous electroweak symmetry breaking, play a crucial role in explaining the origin of the masses of fermions.
The constraints on the Yukawa couplings with the top quark \cite{CMS:2019art,CMS:2020djy}, the bottom quark \cite{ATLAS:2018kot,CMS:2018nsn}, the charm quark \cite{ATLAS:2022ers,CMS:2022psv}, the tau lepton \cite{ATLAS:2022yrq,CMS:2022kdi}, and the muon \cite{ATLAS:2020fzp,CMS:2020xwi} have been set by the ATLAS and CMS collaborations.
Because the top quark is very heavy,
the Higgs boson can not decay into top quarks.
The dominant decay mode of the Higgs boson is thus $H\to b\bar{b}$.
From the signal strength observed at the LHC \cite{ATLAS:2018kot,CMS:2018nsn},
the bottom quark Yukawa coupling is derived with an error of $\pm 10\%$.
The accuracy will be significantly improved at future lepton colliders such as the Circular Electron Positron Collider (CEPC) \cite{An:2018dwb,CEPCPhysicsStudyGroup:2022uwl, Zhu:2022lzv}, and the precision can reach the per mill level \cite{CEPCPhysicsStudyGroup:2022uwl,Zhu:2022lzv}.

The corresponding theoretical predictions have been improved continuously in the past half-century.
The next-to-leading order (NLO) QCD corrections \cite{Braaten:1980yq,Sakai:1980fa,Janot:1989jf,Drees:1990dq} were computed more than 40 years ago. 
And the NLO electroweak (EW) corrections have also been known for a long time \cite{Dabelstein:1991ky,Kniehl:1991ze}. The mixed QCD and EW corrections of order $\mathcal{O}(\alpha\alpha_s)$ have been obtained in Refs. \cite{Kataev:1997cq,Mihaila:2015lwa}. 
Later, the decay width of $H\rightarrow b\bar{b}$ was calculated up to $\mathcal{O}(y_b^2\alpha_s^4)$ \cite{Gorishnii:1990zu,Chetyrkin:1996sr,Baikov:2005rw,Herzog:2017dtz,Chen:2023fba},
when the final-state bottom quark is taken to be massless
but the bottom Yukawa coupling is still finite.
The corresponding differential results were investigated up to $\mathcal{O}(\alpha_s^3)$ \cite{Anastasiou:2011qx,DelDuca:2015zqa,Mondini:2019gid}.
The inclusive and differential decay widths with finite $m_b$ were computed at NNLO numerically \cite{Bernreuther:2018ynm,Behring:2019oci,Somogyi:2020mmk}. 
The effects of parton shower in this decay process have been explored in \cite{Bizon:2019tfo,Hu:2021rkt}.
The differential decay rates of $H\to b\bar{b}j$ at NNLO and $H\to b\bar{b}b\bar{b}$ at NLO were presented in \cite{Mondini:2019vub} and \cite{Gao:2019ypl}, respectively.

The decay $H\to b\bar{b}$ can also be induced by a top quark loop with the top quark Yukawa coupling $y_t$, 
which is much larger than $y_b$.
However, this contribution starts at the  two-loop level, i.e., $\mathcal{O}(y_by_t \alpha_s^2)$.
Because the helicity should be flipped along the bottom quark current,
an additional bottom quark mass suppression emerges.
As a result, the contribution is not significantly larger, or even smaller, than that of $\mathcal{O}(y_b^2\alpha_s^2)$.
The analytic result of the decay width of $H\to b\bar{b}$ at $\mathcal{O}(y_by_t\alpha_s^2)$ with finite $m_b$ and $m_t$ has been calculated in \cite{Primo:2018zby}. The calculation of the differential decay width that is sensitive to $y_t$  was presented up to $\mathcal{O}(\alpha_s^3)$ in \cite{Mondini:2020uyy,Chen:2023fba}.
The decay rate of $H\to $ hadrons including contributions from both $H\to b\bar{b}$ and $H\to gg$ has been computed up to $\mathcal{O}(\alpha_s^4)$ \cite{Chetyrkin:1997vj,Davies:2017xsp}.

In this paper, we provide an exact result for the decay $H\to b\bar{b}$ at $\mathcal{O}(y_b^2\alpha_s^2)$ with full dependence on $m_b$.
Specifically, we consider the contributions from the decay processes $H\rightarrow b\bar{b}(+X),X=g,gg,q\bar{q},b\bar{b}$, where $q$ stands for the massless quark.
We note that an analytical calculation has been performed in \cite{Chetyrkin:1995pd,Harlander:1997xa},
but only the expansion result in $m_b^2/m_H^2$ was obtained.
Our analytic results enable an efficient computation of the NNLO decay width, and can be readily extended to other similar processes, such as a new heavy scalar decaying to top quarks.
Moreover, given the analytic result, we can explore different aspects of the decay rate.
The expansion of the decay width near the threshold shows power divergences, which violate the perturbative expansion in $\alpha_s$.
To understand the origin of these divergences and to restore the power of perturbative expansion, we need to calculate a new Coulomb Green function.

This paper is organized as follows.
Our calculation framework is introduced in section \ref{sec:framework}.
The master integrals  in the calculation of the decay width are computed in section \ref{sec:MIs}.
In section \ref{sec:renorm}, the renormalization procedure is briefly described.
The analytical and numerical results for the decay width are presented in section \ref{sec:analytic} and section \ref{sec:numerical}, respectively.
Our conclusion is given in section \ref{sec:conclusion}.

\section{Calculation framework}
\label{sec:framework}

We are going to calculate the decay width of $H\rightarrow b\bar{b}$, denoted by $\Gamma_{Hb{\bar b}}$,  via the optical theorem,
\begin{align}
\Gamma_{Hb{\bar b}} = \frac{\text{Im}(\Sigma)}{m_H},
\end{align}
where $\Sigma$ represents the amplitude of the process $H \rightarrow b \bar{b} \rightarrow H$.
In this method, we do not need to calculate the virtual and real corrections separately, and the complicated multi-body phase space integration can be avoided. 

We define the contribution from the final states of two bottom quarks as $\tilde{\Gamma}^{y_by_b}_{Hb\bar{b}}$, and that from final states of four bottom quarks as ${\Gamma}^{y_by_b}_{Hb\bar{b}b\bar{b}}$.
The partial decay width of $H\to b\bar{b}$ is thus given by
\begin{align}
	{\Gamma}^{y_by_b}_{Hb\bar{b}}  \equiv \tilde{\Gamma}^{y_by_b}_{Hb\bar{b}} + {\Gamma}^{y_by_b}_{Hb\bar{b}b\bar{b}}\,.
\label{eq:Hbb} 
\end{align}

\begin{figure}[ht]
	\centering
	\begin{minipage}{0.3\linewidth}
		\centering
		\includegraphics[width=0.8\linewidth]{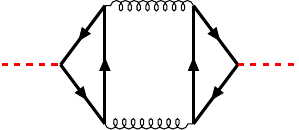}
	\end{minipage}
	\begin{minipage}{0.3\linewidth}
		\centering
		\includegraphics[width=0.8\linewidth]{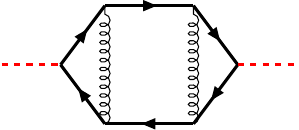}
	\end{minipage}
	\begin{minipage}{0.3\linewidth}
	\centering
	\includegraphics[width=0.8\linewidth]{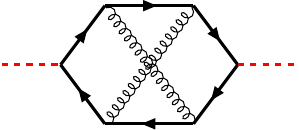}
	\end{minipage}
\caption{Typical three-loop Feynman diagrams of $H \rightarrow b \bar{b} \rightarrow H$. The thick black and red lines stand for the massive bottom quark and the Higgs boson, respectively.}
\label{ThreeLoop}
\end{figure}

\begin{figure}[ht]
	\centering
\begin{minipage}{0.3\linewidth}
	\centering
	\includegraphics[width=0.8\linewidth]{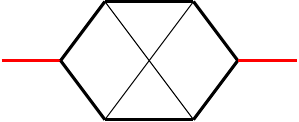}
	\caption*{NP1}
\end{minipage}
\begin{minipage}{0.3\linewidth}
	\centering
	\includegraphics[width=0.8\linewidth]{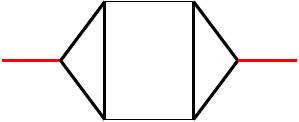}
	\caption*{P1}
\end{minipage}
\caption{The topologies of NP1 and P1 integral families. The thick black and red lines stand for the massive bottom quark and the Higgs boson, respectively. The other lines represent massless particles.}
\label{Fam}
\end{figure}

At  $\mathcal{O}(y_b^2\alpha_s^2)$, 
we have to compute the imaginary part of three-loop self-energy diagrams, as shown in figure~\ref{ThreeLoop}.
We generate the diagrams with the package {\tt FeynArts} \cite{Hahn:2000kx},
and perform the Dirac algebra using the package {\tt FeynCalc} \cite{Shtabovenko:2020gxv}.
The amplitudes are expressed as linear combinations of three-loop scalar integrals, which are reduced to a minimal set of integrals called
master integrals (MIs) due to the identities from
integration by parts (IBP) \cite{Tkachov:1981wb,Chetyrkin:1981qh}.
The solution of these identities is found by the Laporta algorithm \cite{Laporta:2000dsw} that is implemented in the packages {\tt FIRE} \cite{Smirnov:2019qkx} and {\tt Kira} \cite{Klappert:2020nbg}. 
These MIs can be represented by two integral families, i.e., NP1 and P1  in figure~\ref{Fam}. 
In this procedure the package {\tt CalcLoop}\footnote{\url{https://gitlab.com/multiloop-pku/calcloop}} has been used to shift the loop momenta. 
In our calculation, the Higgs boson and the bottom quark are taken to be massive 
while the light quarks are massless.
We do not consider the top quark contribution.

\begin{figure}[ht]
	\centering
	\includegraphics[width=0.6\linewidth]{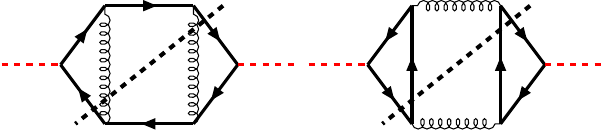}
	\caption{Typical cut diagrams. The black dashed line stands for the cut line, and the propagators crossing the cut line are on-shell. The cut line passes through two (left graph) and four (right graph) bottom quarks.}
	\label{Hbbgg}
\end{figure}

\begin{figure}[ht]
	\centering
	\includegraphics[width=0.13\linewidth]{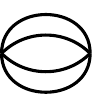}
	\caption{A three-loop massive vacuum bubble diagram.}
	\label{bub}
\end{figure}

According to the cutting rules in the optical theorem, the imaginary parts of $\Sigma$ come from the cut diagrams in which some propagators are put on-shell simultaneously. We find that most cut diagrams in $\Sigma$ have at least two on-shell bottom-quark propagators. 
And these cut diagrams correspond to the processes $H\rightarrow b\bar{b}$, $H\rightarrow b\bar{b}g$, $H\rightarrow b\bar{b}gg$, $H\rightarrow b\bar{b}q\bar{q}$ and $H\rightarrow b\bar{b}b\bar{b}$. 
All of them contribute to $\Gamma_{Hbb}$ at $\mathcal{O}(y_b^2\alpha_s^2)$. 
Two sample cut Feynman diagrams are shown 
in figure~\ref{Hbbgg}.
The requirement of cuts on two bottom quarks simplifies the calculations since some MIs do not have imaginary parts, e.g., the three-loop massive vacuum bubble diagram shown in figure~\ref{bub}.

\begin{figure}[ht]
\centering
\includegraphics[width=0.3\linewidth]{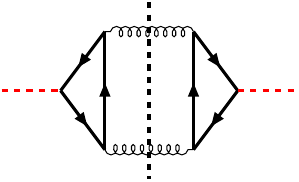}
\caption{A cut diagram contributing to the $H\to gg$ decay.}
\label{Hgg}
\end{figure}

Notice that some cut diagrams 
contribute to the imaginary part of $\Sigma$, but do not have two bottom-quark cut propagators,
such as the diagram of a cut through two gluon propagators as depicted in figure~\ref{Hgg}.
This cut diagram is related to the process $H\rightarrow gg$, which should not be included in the corrections of $H\rightarrow b\bar{b}$.
To subtract this contribution from the imaginary part of the forward scattering amplitude, we calculate the  squared amplitude of $H\rightarrow gg$ with one triangle bottom-quark loop,
and perform the two-body phase space integration.

\section{Calculation of master integrals}
\label{sec:MIs}

As mentioned in eq.(\ref{eq:Hbb}), 
we have to calculate the contributions from both two and four bottom quark final states.
The corresponding MIs can be computed separately.

\begin{figure}[ht]
	\centering
	\begin{minipage}{0.3\linewidth}
		\centering
		\includegraphics[width=0.8\linewidth]{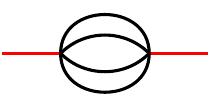}
	\end{minipage}
	\begin{minipage}{0.3\linewidth}
		\centering
		\includegraphics[width=0.8\linewidth]{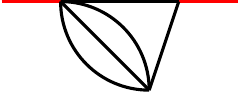}
	\end{minipage}
	\caption{Two master integrals contributing to the decay to four bottom-quark final states. The thick black and red lines stand for the massive bottom quark and the Higgs boson, respectively.}
	\label{ellip}
\end{figure}

The MIs contributing to the decay to four bottom quark final state contain elliptic integrals, such as those shown in figure~\ref{ellip}.
They have been calculated in \cite{Lee:2019wwn} where the authors consider the total cross section of $e^+e^-\rightarrow Q\bar{Q}Q\bar{Q}$.
After choosing a regular basis, only the $\mathcal{O}(\epsilon^0)$ parts of the MIs are needed,
which in turn can be expressed either as complete elliptic integrals of the first kind or one-fold integrals of them.
Following this method, we present the decay width ${\Gamma}^{y_by_b}_{Hb\bar{b}b\bar{b}}$ in terms of one-fold integrals over expressions depending on complete elliptic integrals and multiple polylogarithms.
The explicit form can be found in eq. (\ref{eq:Hto4b}) below.

\begin{figure}[ht]
	\centering
	\begin{minipage}{0.3\linewidth}
		\centering
		\includegraphics[width=0.8\linewidth]{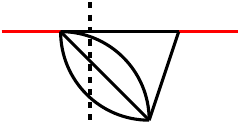}
	\end{minipage}
	\begin{minipage}{0.3\linewidth}
		\centering
		\includegraphics[width=0.8\linewidth]{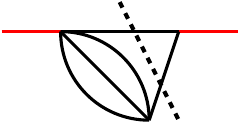}
	\end{minipage}
	\caption{The four (left) and two (right) bottom quark cut diagrams. The thick black and red lines stand for the massive bottom quark and the Higgs boson, respectively.}
	\label{ellipCut}
\end{figure}

Then we turn to the MIs that are needed in the calculation of $\tilde{\Gamma}^{y_by_b}_{Hb\bar{b}}$.
We adopt the method of differential equations \cite{Kotikov:1990kg,Kotikov:1991pm}.
Once a set of canonical basis $\boldsymbol{F}$ is found, the dimensional regulator $\epsilon$ is factorized out from the kinematic variables \cite{Henn:2013pwa},
\begin{align}
d \boldsymbol{F} (\epsilon,x_n) =\epsilon \, d A(x_n) \boldsymbol{F}(\epsilon,x_n),
\end{align}
where $\{x_n\}$ denote the kinematic variables.
When $d A(x_n)$ is expressed in $d\log$ forms, 
the differential equations can be solved recursively in terms of multiple polylogarithms (MPLs) \cite{Goncharov:1998kja}, which are defined by $G(x)\equiv 1$ and
\bqa
	G(l_1,l_2,\ldots,l_n,x) &\equiv & \int_0^x \frac{\text{d} t}{t - l_1} G(l_2,\ldots,l_n,t)\, ,\\
	G(\overrightarrow{0}_n , x) & \equiv & \frac{1}{n!}\ln^n x\, .
\eqa
The number of elements in the set $\{l_1,l_2,\ldots,l_n\}$ is referred to as the transcendental $weight$ of the MPLs.
Note that we have dropped MIs that possess an imaginary part only from cuts on four bottom quarks.
And all the boundary values are evaluated in the phase space where it is not possible to apply a cut on four bottom quarks.
For instance, the left diagram in figure~\ref{ellip} is dropped in the basis.
The right diagram in figure~\ref{ellip} is included but only the cut through two bottom quarks, as shown in the right diagram of figure~\ref{ellipCut}, is considered.

\subsection{Canonical basis of NP1}
The master integrals of the NP1 integral family in figure \ref{Fam} are defined by 
\begin{align}
	I^{\rm NP1}_{n_1,n_2,\ldots,n_{9}}=\textrm{Im}\int{\mathcal D}^D q_1~{\mathcal D}^D q_2\frac{D_9^{-n_9}}{D_1^{n_1}~D_2^{n_2}~D_3^{n_3}~D_4^{n_4}~D_5^{n_5}~D_6^{n_6}~D_7^{n_7}D_8^{n_8}}
\end{align}
with
\begin{align}
	{\mathcal D}^D q_i = \frac{\left(m_b^2 \right)^\epsilon}{i \pi^{D/2}\Gamma(1+\epsilon)}  d^D q_i \ ,\quad D=4-2\epsilon \,,
\end{align}
where all $n_i\ge 0,i=1,\cdots, 8$, and $n_9\le 0$. The denominators $D_i$ read
\begin{align}
	D_1 &= q_1^2-m_b^2,&
	D_2 &= q_2^2-m_b^2,&
	D_3 &= (q_1-q_2)^2,\nonumber\\
	D_4 &= q_3^2-m_b^2,&
	D_5 &= (q_2+q_3)^2,&
	D_6 &= (q_3-k)^2-m_b^2,\nonumber\\
	D_7 &= (q_1+k)^2-m_b^2,&
	D_8 &=(q_1-q_2-q_3+k)^2-m_b^2,&
	D_9 &=(q_1-q_3+k)^2,
\end{align}
where the Feynman prescription $+i\varepsilon$ has been suppressed. The external momentum $k$ satisfies
$
k^2 = m_H^2.
$
We focus only on the imaginary part of the integral, which gets contributions from all possible cuts except for the ones on four bottom quarks.
The results of the master integrals for specific cuts have been obtained in \cite{Lee:2020mvt}.

In the NP1 integral family, there are 29 MIs contributing to $\tilde{\Gamma}_{Hb\bar{b}}$. 
To construct the canonical basis, we first select the MIs 
\begin{align}
\M^{\text{NP1}}_1&=\epsilon ^3 m_b^2I^{\text{NP1}}_{2,2,0,2,0,1,0,0,0},\quad&
\M^{\text{NP1}}_2&=\epsilon ^3 m_b^2I^{\text{NP1}}_{2,2,0,0,1,2,0,0,0},\quad\nonumber\\
\M^{\text{NP1}}_3&=\epsilon ^3 m_b^2I^{\text{NP1}}_{2,2,0,0,2,1,0,0,0},\quad&
\M^{\text{NP1}}_4&=\epsilon ^3 m_b^2I^{\text{NP1}}_{2,0,2,0,1,2,0,0,0},\quad\nonumber\\
\M^{\text{NP1}}_5&=\epsilon ^3 m_b^2I^{\text{NP1}}_{2,0,2,0,2,1,0,0,0},\quad&
\M^{\text{NP1}}_6&=(1-2 \epsilon ) \epsilon  m_b^4I^{\text{NP1}}_{3,0,1,0,1,3,0,0,0},\quad\nonumber\\
\M^{\text{NP1}}_7&=\epsilon ^3 m_b^4I^{\text{NP1}}_{2,2,0,2,0,1,1,0,0},\quad&
\M^{\text{NP1}}_8&=\epsilon ^3 m_b^4I^{\text{NP1}}_{0,2,1,2,0,1,2,0,0},\quad\nonumber\\
\M^{\text{NP1}}_9&=\epsilon ^3 m_b^4I^{\text{NP1}}_{0,2,2,2,0,1,1,0,0},\quad&
\M^{\text{NP1}}_{10}&=\epsilon ^3 m_b^2I^{\text{NP1}}_{0,1,1,0,1,2,2,0,0},\quad\nonumber\\
\M^{\text{NP1}}_{11}&=\epsilon ^2 m_b^4I^{\text{NP1}}_{0,1,1,0,1,3,2,0,0},\quad&
\M^{\text{NP1}}_{12}&=(1-2 \epsilon ) \epsilon ^3 m_b^2I^{\text{NP1}}_{0,2,1,0,1,2,1,0,0},\quad\nonumber\\
\M^{\text{NP1}}_{13}&=(1-2 \epsilon ) \epsilon ^3 m_b^2I^{\text{NP1}}_{2,2,0,0,1,1,0,1,0},\quad&
\M^{\text{NP1}}_{14}&=\epsilon ^3 m_b^4I^{\text{NP1}}_{2,2,0,0,1,2,0,1,0},\quad\nonumber\\
\M^{\text{NP1}}_{15}&=(1-2 \epsilon ) \epsilon ^3 m_b^2I^{\text{NP1}}_{2,0,2,1,1,1,0,0,0},\quad&
\M^{\text{NP1}}_{16}&=(1-2 \epsilon ) \epsilon ^3 m_b^2I^{\text{NP1}}_{2,2,0,1,0,1,0,1,0},\quad\nonumber\\
\M^{\text{NP1}}_{17}&=(1-2 \epsilon ) \epsilon ^4 m_b^2I^{\text{NP1}}_{1,0,2,1,1,1,1,0,0},\quad&
\M^{\text{NP1}}_{18}&=(1-2 \epsilon ) \epsilon ^4 m_b^2I^{\text{NP1}}_{1,1,1,0,1,2,1,0,0},\quad\nonumber\\
\M^{\text{NP1}}_{19}&=(1-2 \epsilon ) \epsilon ^4 m_b^2I^{\text{NP1}}_{2,1,0,1,1,1,0,1,0},\quad&
\M^{\text{NP1}}_{20}&=(1-2 \epsilon ) \epsilon ^3 m_b^4I^{\text{NP1}}_{2,1,0,1,1,1,0,2,0},\quad\nonumber\\
\M^{\text{NP1}}_{21}&=(1-2 \epsilon ) \epsilon ^5 m_b^2I^{\text{NP1}}_{1,1,1,1,1,1,0,1,0},\quad&
\M^{\text{NP1}}_{22}&=(1-2 \epsilon )^2 \epsilon ^4 m_b^2I^{\text{NP1}}_{1,1,1,0,0,1,1,1,1},\quad\nonumber\\
\M^{\text{NP1}}_{23}&=(1-2 \epsilon ) \epsilon ^4 m_b^4I^{\text{NP1}}_{1,1,1,0,0,1,2,1,1},\quad&
\M^{\text{NP1}}_{24}&=(1-2 \epsilon ) \epsilon ^4 m_b^2I^{\text{NP1}}_{1,2,0,1,0,1,1,1,0},\quad\nonumber\\
\M^{\text{NP1}}_{25}&=(1-2 \epsilon ) \epsilon ^3 m_b^4I^{\text{NP1}}_{2,2,0,1,0,1,1,1,0},\quad&
\M^{\text{NP1}}_{26}&=(1-2 \epsilon ) \epsilon ^4 m_b^4I^{\text{NP1}}_{1,2,1,1,1,1,1,0,0},\quad\nonumber\\
\M^{\text{NP1}}_{27}&=(1-2 \epsilon ) \epsilon ^4 m_b^4I^{\text{NP1}}_{2,1,1,1,1,1,1,0,0},\quad&
\M^{\text{NP1}}_{28}&=(1-2 \epsilon ) \epsilon ^5 m_b^4I^{\text{NP1}}_{1,1,1,1,1,1,1,1,0},\quad\nonumber\\
\M^{\text{NP1}}_{29}&=(1-2 \epsilon ) \epsilon ^5 m_b^2I^{\text{NP1}}_{1,1,1,1,1,1,1,1,-1}\,.
\end{align}
Here all $\M^{\text{NP1}}_{i}$ are dimensionless, and the corresponding topology diagrams are displayed in figure \ref{NP1_Topo} in appendix \ref{appendix:topoMIs}.

Then the canonical basis integrals $F^\text{NP1}_{i},~i=1,\dots,29,$ can be constructed as linear combinations of $\M^\text{NP1}_{i}$ using a method similar to that in ref. \cite{Argeri:2014qva}. 
Examples can be found in \cite{Chen:2021gjv}.
For simplicity, we define a dimensionless variable $z \equiv m_H^2/m_b^2+i0^+$. Then we obtain the following canonical basis of the NP1 family: 
\begin{align}
	F^{\text{NP1}}_1&=\M^{\text{NP1}}_1 r,\quad F^{\text{NP1}}_2=\M^{\text{NP1}}_2 \left(-z\right), \quad F^{\text{NP1}}_3=\frac{\M^{\text{NP1}}_2 r}{2}+\M^{\text{NP1}}_3 r, \quad
	\nonumber\\
	F^{\text{NP1}}_4&=\M^{\text{NP1}}_4 \left(-z\right),\quad
	F^{\text{NP1}}_5=\left(\M^{\text{NP1}}_4+\M^{\text{NP1}}_5\right) (-r),\quad F^{\text{NP1}}_6=-\frac{\M^{\text{NP1}}_5 \left(4-z\right)}{4}+\M^{\text{NP1}}_6,\quad
	\nonumber\\
	F^{\text{NP1}}_7&=\M^{\text{NP1}}_7 z \left(z-4\right),\quad 
	F^{\text{NP1}}_8=\M^{\text{NP1}}_8 r z,\quad F^{\text{NP1}}_9=\frac{\M^{\text{NP1}}_8 \left(z-4\right) z}{2}+\M^{\text{NP1}}_9 \left(z-4\right) z,
	\nonumber\\
	F^{\text{NP1}}_{10}&=\frac{\M^{\text{NP1}}_2 \left(4-3 z\right)}{8-4 z}+\frac{ \M^{\text{NP1}}_4 z}{4}+\frac{\M^{\text{NP1}}_{10} \left(2 \epsilon  \left(4 z^2-9 z-4\right)+4 z^2-11 z+4\right)}{8-4 z}
	\nonumber\\&\quad
	-\frac{2 \M^{\text{NP1}}_{11} \epsilon  \left(2 z-5\right) z}{z-2}+\frac{\M^{\text{NP1}}_{12} \left(z-4\right) \left(z-1\right)}{4 \left(z-2\right)},\quad 
	\nonumber\\
	F^{\text{NP1}}_{11}&=\frac{\M^{\text{NP1}}_2 r \left(z^2-6 z+4\right)}{8 \left(z-4\right)\left(z-2\right)}+\frac{\M^{\text{NP1}}_{10} r \left(2 \epsilon  \left(z^2+2 z-11\right)+z^2-4 z+3\right)}{2 \left(z-4\right) \left(z-2\right)}
	\nonumber\\&\quad
	+\frac{\M^{\text{NP1}}_{11} r \left(2 \epsilon  \left(z^2-6\right)-z+2\right)}{\left(z-4\right) \left(z-2\right)}
	-\frac{\M^{\text{NP1}}_{12} r \left(z-3\right) \left(z-1\right)}{2 \left(z-4\right) \left(z-2\right)},\quad 
	\nonumber\\
	F^{\text{NP1}}_{12}&=-\frac{\M^{\text{NP1}}_2 z}{z-2}+\M^{\text{NP1}}_4 z+\frac{\M^{\text{NP1}}_{10} \left(6 \epsilon  z+z\right)}{2-z}-\frac{8 \M^{\text{NP1}}_{11} \epsilon  z}{z-2}+\frac{\M^{\text{NP1}}_{12} \left(z-1\right) z}{z-2},
	\nonumber\\
	F^{\text{NP1}}_{13}&=\M^{\text{NP1}}_{13} z+\M^{\text{NP1}}_{14} z^2,\quad  F^{\text{NP1}}_{14}=\M^{\text{NP1}}_{14} \left(-z\right),\quad 
	\nonumber\\
	F^{\text{NP1}}_{15}&=-\frac{\M^{\text{NP1}}_4 r \left(7 z-12\right)}{12 \left(4-z\right)}+\frac{\M^{\text{NP1}}_5 r}{12}-\frac{\M^{\text{NP1}}_6 r}{3 \left(4-z\right)}+\frac{\M^{\text{NP1}}_{15} r}{4-z},\quad F^{\text{NP1}}_{16}=-\frac{\M^{\text{NP1}}_{16} r}{4-z},\quad 
	\nonumber\\
	F^{\text{NP1}}_{17}&=\M^{\text{NP1}}_{17} \left(-z\right), \quad F^{\text{NP1}}_{18}=\M^{\text{NP1}}_{18} \left(-z\right),\quad F^{\text{NP1}}_{19}=\M^{\text{NP1}}_{19} \left(-z\right),
	\nonumber\\
	F^{\text{NP1}}_{20}&=-2 \M^{\text{NP1}}_{14} r z-\frac{3 \M^{\text{NP1}}_{16} r z}{z-4}+\M^{\text{NP1}}_{20} r z,\quad F^{\text{NP1}}_{21}=\M^{\text{NP1}}_{21} z,
	\nonumber\\
	F^{\text{NP1}}_{22}&=\frac{1}{4} z \left(\M^{\text{NP1}}_{22}+\M^{\text{NP1}}_{23} \left(z-4\right)\right),\quad
	\nonumber\\
	F^{\text{NP1}}_{23}&=-\M^{\text{NP1}}_{18} r+2 \M^{\text{NP1}}_{19} r+2 \M^{\text{NP1}}_{21} r+\M^{\text{NP1}}_{22} r+\M^{\text{NP1}}_{23} r z,\quad
	\nonumber\\
	F^{\text{NP1}}_{24}&=\M^{\text{NP1}}_{24} \left(-z\right),\quad F^{\text{NP1}}_{25}=-\frac{\M^{\text{NP1}}_{16} r z}{z-4}+\M^{\text{NP1}}_{25} r z,\quad F^{\text{NP1}}_{26}=\M^{\text{NP1}}_{26} \left(-z\right),\quad
	\nonumber\\
	F^{\text{NP1}}_{27}&=\frac{3 \M^{\text{NP1}}_2 r z^2}{8 \left(z-2\right)^2}+\frac{3 \M^{\text{NP1}}_4 r z}{8 \left(z-2\right)}-\frac{3 \M^{\text{NP1}}_8 r \left(z-4\right) z}{8 \left(z-2\right)}-\frac{3 \M^{\text{NP1}}_9 r \left(z-4\right) z}{4 \left(z-2\right)}
	\nonumber\\&\quad
	+\frac{3 \M^{\text{NP1}}_{10} r (6 \epsilon +1) z}{4 \left(z-2\right)^2}+\frac{6 \M^{\text{NP1}}_{11} r \epsilon  z}{\left(z-2\right)^2}-\frac{3 \M^{\text{NP1}}_{12} r \left(z-1\right) z}{4 \left(z-2\right)^2}
	+\frac{3 \M^{\text{NP1}}_{17} r z}{2 \left(z-2\right)}
	\nonumber\\&\quad
	+\frac{\M^{\text{NP1}}_{18} r z}{4 \left(z-2\right)}+\frac{\M^{\text{NP1}}_{26} r z}{z-2}-\frac{\M^{\text{NP1}}_{27} r z}{z-2},\quad
	\nonumber\\
	F^{\text{NP1}}_{28}&=\M^{\text{NP1}}_{28} (-r) z,\quad F^{\text{NP1}}_{29}=\M^{\text{NP1}}_{28} \left(z-4\right) z+\M^{\text{NP1}}_{29} z,
\end{align}
where
\begin{align}
r = \sqrt{z(z-4)}.
\end{align}
To rationalize the square root $r$, we write
\begin{align}
	z = -\frac{(w-1)^2}{w}
\end{align}
with  $-1<w<0$.
Then $r$ is rationalized to 
\begin{align}
r = \frac{(w+1) (w-1)}{w}.
\end{align}
The corresponding differential equations for the canonical basis $\boldsymbol{F}^{\text{NP1}}$ become 
\begin{align}
\frac{d \boldsymbol{F}^{\text{NP1}}}{dw} = \epsilon \left( \sum_{i=1}^5 \frac{\boldsymbol{N_i}^{\text{NP1}}}{w-l_i^{\text{NP1}}}\right) \boldsymbol{F}^{\text{NP1}}
\end{align}
with
\begin{align}
l^{\text{NP1}}_1 = 0\,,~ 
l^{\text{NP1}}_{2} = 1\,,~
l^{\text{NP1}}_{3} = -1\,,~
l^{\text{NP1}}_{4} = \frac{1+\sqrt{3} i}{2}\,,~
l^{\text{NP1}}_{5} = \frac{1-\sqrt{3} i}{2}
\end{align}
and $\boldsymbol{N_i}^{\text{NP1}}$ being constant rational matrices.

\subsection{Canonical basis of P1}
The P1 integral family  is defined by 
\begin{align}
	I^{P1}_{n_1,n_2,\ldots,n_{9}}=\textrm{Im}\int{\mathcal D}^D q_1~{\mathcal D}^D q_2\frac{D_9^{-n_9}}{D_1^{n_1}~D_2^{n_2}~D_3^{n_3}~D_4^{n_4}~D_5^{n_5}~D_6^{n_6}~D_7^{n_7}D_8^{n_8}},
\end{align}
where the denominators $D_i$ read
\begin{align}
	D_1 &= q_1^2-m_b^2,&
	D_2 &= (q_1+k_1)^2-m_b^2,&
	D_3 &= q_3^2-m_b^2,\nonumber\\
	D_4 &= (q_3^2-k_1)^2-m_b^2,&
	D_5 &= (q_1-q_2)^2-m_b^2,&
	D_6 &= q_2^2,\nonumber\\
	D_7 &= (q_2+q_3)^2-m_b^2,&
	D_8 &=(q_2+k_1)^2,&
	D_9 &=(q_1+q_3)^2.
\end{align}
There are 19 MIs in this integral family contributing to $\tilde{\Gamma}_{Hb\bar{b}}$. 
Most of them have appeared in the NP1 integral family discussed above. 
Here we only show the new MIs,
\begin{align}
\M^{\text{P1}}_1&=\epsilon ^3 m_b^2
I^{\text{P1}}_{2,0,2,0,0,2,0,1,0},\quad&
\M^{\text{P1}}_5&=\epsilon ^3 m_b^4
I^{\text{P1}}_{2,1,2,0,0,2,0,1,0},\quad\nonumber\\
\M^{\text{P1}}_{12}&=\epsilon ^3 m_b^6
I^{\text{P1}}_{2,1,2,1,0,2,0,1,0},\quad&
\M^{\text{P1}}_{13}&=(1-2\epsilon)\epsilon^4 m_b^2
I^{\text{P1}}_{1,1,2,0,1,1,0,1,0},\quad&\nonumber\\
\M^{\text{P1}}_{18}&=(1-2\epsilon)\epsilon^4 m_b^4
I^{\text{P1}}_{1,1,2,1,1,1,0,1,0},\quad&
\M^{\text{P1}}_{19}&=(1-2\epsilon)\epsilon^5 m_b^4
I^{\text{P1}}_{1,1,1,1,1,1,1,1,0}.\quad&
\end{align}
The topology diagrams are displayed in figure \ref{P1_Topo} in appendix \ref{appendix:topoMIs}. 
We find the corresponding canonical basis integrals,
\begin{align}
F^{\text{P1}}_1&=\M^{\text{P1}}_1 \left(-z\right), \quad
F^{\text{P1}}_5=\M^{\text{P1}}_5 (-r) z, \quad
F^{\text{P1}}_{12}=-\M^{\text{P1}}_{12} z^2 \left(z-4\right),\quad
\nonumber\\
F^{\text{P1}}_{13}&=\M^{\text{P1}}_{13} \left(-z\right), \quad
F^{\text{P1}}_{18}=\M^{\text{P1}}_{18} (-r) z, \quad F^{\text{P1}}_{19}=\M^{\text{P1}}_{19} z^2.
\end{align}
The corresponding differential equations are 
\begin{align}
\frac{d \boldsymbol{F}^{\text{P1}}}{dw} = \epsilon \left( \sum_{i=1}^3 \frac{\boldsymbol{N_i}^{\text{P1}}}{w-l_i^{\text{P1}}}\right) \boldsymbol{F}^{\text{P1}}
\end{align}
with
\begin{align}
l^{\text{P1}}_1 = 0\,,~ 
l^{\text{P1}}_{2} = 1\,,~
l^{\text{P1}}_{3} = -1\,,~
\end{align}
and $\boldsymbol{N_i}^{\text{P1}}$ being constant rational matrices.

\subsection{Boundary conditions}

Given the above canonical differential equations, we obtain the results of the basis integrals at each order of $\epsilon$ up to some integration constants.
The MPLs in the results can be evaluated efficiently using the {\tt PolyLogTools}  package \cite{Duhr:2019tlz}, which
depends on the algorithm of ref. \cite{Vollinga:2004sn} in the {\tt GiNaC}  \cite{Bauer:2000cp} framework.
The integration constants are calculated numerically by comparing the results of basis integrals  with those obtained using the  {\tt AMFlow} package \cite{Liu:2017jxz,Liu:2020kpc,Liu:2022chg,Liu:2022mfb}.
Their analytical form is reconstructed using the PSLQ algorithm \cite{Ferguson1992,Ferguson:1999aa}.
We find that the transcendental constants relevant to our calculation consist of 
\begin{align}
\{\log(2), \pi, \zeta(3)\}
\end{align} 
up to weight four,
where $\zeta(n)$ is the Riemann zeta function.
All the master integrals are then checked against the numerical computation with {\tt AMFlow},
and over 200-digit agreement is achieved.
We provide the analytical results of the canonical bases in an ancillary file along with the arXiv submission.

\section{Renormalization}
\label{sec:renorm}

In the framework of the optical theorem, 
the imaginary part of each three-loop self energy diagram arises from all possible cuts,
corresponding to the sum of both virtual and real corrections.
The infrared divergences cancel out in the sum, and only ultraviolet divergences remain.

We perform renormalization following the standard procedure, i.e.,
by including the contribution from diagrams with counter-terms.
The bottom quark mass is renormalized in the on-shell scheme,
while the strong coupling $\als$ is renormalized in the $\overline{\rm MS}$ scheme,
\begin{align}
m_{b,bare} &= Z_m m_b,\nonumber\\
\als^{bare}C^{\eps} &= \mu^{2\eps} \als Z_{\als},    
\end{align}
where $ C_{\eps} = (4\pi e^{- \gamma_E} )^{\eps} $.
The renormalization constants up to $\als^2$ are given by \cite{Caswell:1974gg,Jones:1974mm,Czakon:2007wk,Melnikov:2000qh},
\begin{align}
Z_{\als} &= 1+ \left(\frac{\als}{4\pi}\right)\frac{2n_l-31}{3\eps} +  \left(\frac{\als}{4\pi}\right)^2\left(\frac{\left(31-2 n_l\right)^2}{9 \epsilon ^2}+\frac{19 n_l-134}{3\epsilon }\right),\nonumber\\
Z^{\overline{\rm MS}}_m &= 1 - \left(\frac{\als}{4\pi}\right)\frac{4}{\eps} +\left(\frac{\als}{4\pi}\right)^2\left(\frac{2 (43-2n_l)}{3 \eps^2}+\frac{10n_l-293}{9\eps}\right),\nonumber\\
Z^{\text{OS}}_m &= 1 +\left(\frac{\als}{4\pi}\right)D_{\eps}\left(-\frac{4}{\epsilon }-\frac{16}{3}-\frac{32 \epsilon }{3}-\frac{64 \epsilon ^2}{3}\right)\nonumber\\
&\quad+\left(\frac{\als}{4\pi}\right)^2D_{\eps}^2\bigg(\frac{86-4 n_l}{3\eps^2}+\frac{10 n_l-101}{9 \epsilon }+\frac{4 \left(2 n_l-31\right)}{3 \epsilon}\log \left(\frac{\mu^2}{m^2}\right)\nonumber\\
&\quad+\frac{2(31-2 n_l)}{3}\log^2\left(\frac{\mu^2}{m^2}\right) +\frac{16(2 n_l-31)}{9}\log\left(\frac{\mu^2}{m^2}\right)
+\frac{\left(10 n_l-79\right)\pi ^2}{9}\nonumber\\
&\quad
+\frac{ \left(142 n_l-2251\right)}{18}+\frac{8 \zeta (3)}{3}-\frac{16\pi ^2 \log (2)}{9}
\bigg),
\end{align}
where
\begin{align}
     D_{\eps} \equiv \frac{ \Gamma(1+\eps)}{e^{-\gamma_E\eps}} \left(\frac{ \mu^2}{m^2}\right)^{\eps}
\end{align}
and color factors $C_F=4/3$, $C_A=3$, $T_R=1/2$ have been substituted.

The bottom quark Yukawa coupling is related to the bottom quark mass.
However, one has the freedom to choose a different renormalization scheme.
Though it is expected that the final result for the decay width should be the same if all order corrections are included,
the fixed-order results in different schemes may exhibit obvious deviations, which should be taken as one kind of theoretical uncertainty.
The results with Yukawa couplings in the on-shell and $\overline{\rm MS}$ scheme can be converted to each other by applying the relations between the two schemes, which have been computed up to $\mathcal{O}(\als^4)$ \cite{Broadhurst:1991fy,Gray:1990yh,Marquard:2015qpa,Marquard:2016dcn}. 
We have performed two separate calculations using different renormalization schemes.
The ultraviolet divergences cancel in either scheme, and the final results agree with each other after the conversion of the renormalization scheme.

\section{Analytic results}
\label{sec:analytic}

\subsection{Full results}

The decay width of $H\rightarrow b\bar{b}$ up to  $\mathcal{O}(\alpha_s^2)$ can be expressed as 
\begin{align}
\Gamma(H\rightarrow b\bar{b}) = \Gamma_{\text{LO}}\left[1+\frac{\als(\mu)}{\pi}X_1^{y_by_b}+\left(\frac{\als(\mu)}{\pi}\right)^2(X_2^{y_by_b}+X_2^{y_by_t})\right]. 
\end{align}
The LO result is
\begin{align}
\Gamma_{\text{LO}} = \frac{N_c}{16\pi}\overline{y}_b^2m_H\left(1-\frac{4}{z}\right)^{3/2}\,,
\end{align}
where $N_c = 3$ is the number of colors
and $\overline{y}_b$ is  the $\overline{\rm MS}$ Yukawa coupling 
\begin{align}
\overline{y}_b(\mu) = \overline{m_b}(\mu)\left(2\sqrt{2}G_F\right)^{1/2}
\end{align}
with $\overline{m_b}$  the $\overline{\rm MS}$ mass of the bottom quark and $G_F$  the Fermi constant.

The NLO coefficient $X_1^{y_by_b}$ reads
\begin{align}
&X_1^{y_by_b}(w) =2 \log\left(\frac{\mu^2}{m_H^2}\right)+\frac{8}{3(w^2-1)}\Big[2\left(- w^2+i \pi  \left(w^2+1\right)+1\right) G(-1,w)
\nonumber\\&\quad
+\frac{1}{2}\left(w^2+2 i \pi  \left(w^2+1\right)-1\right) G(1,w) +\frac{w\left(3 w^3-w^2-8 w-7\right)G(0,w)}{2(w+1)^2}
\nonumber\\&\quad
+\left(w^2+1\right) \left(-2G(-1,0,w)+2G(0,-1,w)+G(0,1,w)-G(1,0,w)\right)
\nonumber\\&\quad
+\frac{(17-12 i \pi ) w^4+(40+4 i \pi ) w^3+32 i \pi  w^2+(-40+28 i \pi ) w-17}{8(w+1)^2}\Big].
\end{align}
Though there are $i\pi$ terms in the expression, the full result is real since the MPLs also contain imaginary parts. 

The $\mathcal{O}(\alpha_s^2)$ corrections include two parts, i.e., $X_2^{y_by_b}$ and $X_2^{y_by_t}$. 
The latter denotes the contribution from the interference of the bottom and top quark Yukawa coupling induced amplitudes and starts at $\mathcal{O}(\alpha_s^2)$.
The corresponding analytical result has been calculated in ref. \cite{Primo:2018zby}.
The former receives contributions from both two and four bottom quark final states, as discussed in section \ref{sec:framework}, 
\begin{align}
X_2^{y_by_b} = \tilde{X}_{2,b\bar{b}}^{y_by_b} + X_{2,b\bar{b}b\bar{b}}^{y_by_b}\,.
\end{align}
The result of  $\tilde{X}_{2,b\bar{b}}^{y_by_b}$ is given by  
\begin{align}
&\tilde{X}_{2,b\bar{b}}^{y_by_b}(w) = -\frac{4 \left(w^2+1\right)^2}{9 \left(w^2-1\right)^2}\Big[100 G(0,-1,-1,0,w)+28 G(0,-1,0,-1,w)
\nonumber\\&\quad
+14 G(0,-1,0,1,w)
+76 G(0,-1,1,0,w)-128 G(0,0,-1,-1,w)
-64 G(0,0,-1,1,w)
\nonumber\\&\quad
-64 G(0,0,1,-1,w)
-32 G(0,0,1,1,w)+76 G(0,1,-1,0,w)-4 G(0,1,0,-1,w)
\nonumber\\&\quad
-2 G(0,1,0,1,w)+35 G(0,1,1,0,w)\Big]
-\frac{32 \left(w^4+10 w^2+1\right)}{9 (w-1) (w+1)^3}\Big[-6 G(1,x_1,0,-1,w)
\nonumber\\&\quad
+4 G(1,x_1,0,0,w)
	-3 G(1,x_1,0,1,w)
	+G(1,x_1,1,0,w)-6 G(1,x_2,0,-1,w)
\nonumber\\&\quad
 +4 G(1,x_2,0,0,w)
	-3 G(1,x_2,0,1,w)+G(1,x_2,1,0,w)+6 G(1,0,0,-1,w)
\nonumber\\&\quad 
 +3 G(1,0,0,1,w)
-G(1,0,1,0,w)\Big]-\frac{32 \left(2 w^5+w^4+8 w^3-2 w^2+2 w+1\right)}{9 (w-1)^2 (w+1)^3}
\nonumber\\&\quad
\times\Big[6 G(0,x_1,0,-1,w)
	-4 G(0,x_1,0,0,w)
	+3 G(0,x_1,0,1,w)-G(0,x_1,1,0,w)
\nonumber\\&\quad
+6 G(0,x_2,0,-1,w)-4 G(0,x_2,0,0,w)+3 G(0,x_2,0,1,w)-G(0,x_2,1,0,w)\Big]
	\nonumber\\&\quad
	+\frac{1}{9 \left(w^2-1\right)^3}\Big[
	-8 \left(w^4-w^3-16 w^2-w+1\right) (w-1)^2 G(1,0,0,0,w)
	\nonumber\\&\quad
	-16 w \left(6 w^4+2 w^3+31 w^2-19 w+4\right) (w-1) G(0,0,0,-1,w)
	\nonumber\\&\quad
	+w \left(6 w^5-30 w^4-113 w^3-194 w^2-119 w-30\right) G(0,0,0,0,w)
	\nonumber\\&\quad
	-4 w \left(12 w^5+w^4+94 w^3-46 w^2+82 w+1\right) G(0,0,0,1,w)
	\nonumber\\&\quad
	+4 \left(90 w^6-96 w^5+741 w^4-1226 w^3+523 w^2-96 w-128\right) G(0,-1,0,0,w)
	\nonumber\\&\quad
	+32 \left(6 w^6+4 w^5-35 w^4+94 w^3-33 w^2+4 w+8\right) G(0,0,-1,0,w)
	\nonumber\\&\quad
	+2 \left(94 w^6+48 w^5-431 w^4+1118 w^3-413 w^2+48 w+112\right) G(0,0,1,0,w)
	\nonumber\\&\quad
	+2 \left(29 w^6-60 w^5+959 w^4-1628 w^3+749 w^2-60 w-181\right) G(0,1,0,0,w)
	\Big]
	\nonumber\\&\quad
	+\frac{\pi ^4 \left(231 w^6-528 w^5-1774 w^4-5226 w^3-1986 w^2-528 w+19\right)}{1620 \left(w^2-1\right)^3}
	\nonumber\\&\quad
	-\frac{4 i \pi ^3 \left(w^4-w^3-16 w^2-w+1\right) G(1,w)}{9 (w-1) (w+1)^3}
	\nonumber\\&\quad
	+\frac{256 \pi ^2 \left(w^4+10 w^2+1\right)(G(1,x_1,w)+G(1,x_2,w))}{27 (w-1) (w+1)^3}
	\nonumber\\&\quad
	-\frac{256 \pi ^2 \left(2 w^5+w^4+8 w^3-2 w^2+2 w+1\right) (G(0,x_1,w)+G(0,x_2,w))}{27 (w-1)^2 (w+1)^3}
	\nonumber\\&\quad
	+\frac{i \pi ^3 w \left(10 w^5-18 w^4-65 w^3-26 w^2-75 w-18\right) G(0,w)}{27 \left(w^2-1\right)^3}
	\nonumber\\&\quad
	+\frac{2 \pi ^2}{27 \left(w^2-1\right)^3} \Big[ 10 \left(w^4-w^3-16 w^2-w+1\right) (w-1)^2 G(1,0,w)
	\nonumber\\&\quad
	-w\left(8 w^5-24 w^4-89 w^3-110 w^2-97w-24\right) G(0,0,w)
	\nonumber\\&\quad
	-\left(361 w^6-384 w^5+2956 w^4-4698 w^3+2084 w^2-384 w-511\right)G(0,-1,w)
	\nonumber\\&\quad
	-3\left(19 w^6-40 w^5+642 w^4-1154 w^3+502 w^2-40 w-121\right)G(0,1,w)\Big]
	\nonumber\\&\quad
	-\frac{32 \left(w^4+10 w^2+1\right) \zeta (3) G(1,w)}{3 (w-1) (w+1)^3}
	\nonumber\\&\quad
	-\frac{2 \left(15 w^6-18 w^5-78 w^4+386 w^3-94 w^2-18 w-1\right) \zeta (3) G(0,w)}{9 \left(w^2-1\right)^3}
	\nonumber\\&\quad
	-\frac{32 i \pi  \left(w^4+10 w^2+1\right)}{9 (w-1) (w+1)^3}\Big[-4 G(1,x_1,0,w)-G(1,x_1,1,w)-4 G(1,x_2,0,w)
	\nonumber\\&\quad
	-G(1,x_2,1,w)+G(1,0,1,w)\Big]-\frac{32 i \pi  \left(2 w^5+w^4+8 w^3-2 w^2+2 w+1\right)}{9 (w-1)^2 (w+1)^3}
 	\nonumber\\&\quad
 \times\Big[4 G(0,x_1,0,w)+G(0,x_1,1,w)+4 G(0,x_2,0,w)+G(0,x_2,1,w)\Big]
 	\nonumber\\&\quad
	+\frac{4 i \pi  \left(w^2+1\right)^2}{9 \left(w^2-1\right)^2}\Big[100 G(0,-1,-1,w)
	+76 G(0,-1,1,w)+76 G(0,1,-1,w)
 	\nonumber\\&\quad
 +35 G(0,1,1,w)\Big]
	+\frac{\pi  i}{9 \left(w^2-1\right)^3}
 \nonumber\\&\quad
 \times\Big[-w \left(6 w^5-30 w^4-113 w^3-194 w^2-119 w-30\right) G(0,0,0,w)
	\nonumber\\&\quad
	-4 \left(90 w^6-96 w^5+741 w^4-1226 w^3+523 w^2-96 w-128\right) G(0,-1,0,w)
	\nonumber\\&\quad
	-32 \left(6 w^6+4 w^5-35 w^4+94 w^3-33 w^2+4 w+8\right) G(0,0,-1,w)
	\nonumber\\&\quad
	-2 \left(94 w^6+48 w^5-431 w^4+1118 w^3-413 w^2+48 w+112\right) G(0,0,1,w)
	\nonumber\\&\quad
	-2 \left(29 w^6-60 w^5+959 w^4-1628 w^3+749 w^2-60 w-181\right) G(0,1,0,w)
	\nonumber\\&\quad
	+2 \left(15 w^6-18 w^5-78 w^4+386 w^3-94 w^2-18 w-1\right)\zeta (3)\Big]
	\nonumber\\&\quad
	+\frac{8 i \pi  \left(w^4-w^3-16 w^2-w+1\right) G(1,0,0,w)}{9 (w-1) (w+1)^3}
 \nonumber\\&\quad
+\frac{2 (2 n_l-55) (w^2+1)}{9 \left(w^2-1\right)}\log \left(\frac{\mu ^2}{m_H^2}\right) \Big[-2 i \pi  G(-1,w)-i \pi  G(1,w)+2 G(-1,0,w)
\nonumber\\&\quad
-2 G(0,-1,w)-G(0,1,w)+G(1,0,w)\Big]+\frac{4 \left(w^2+1\right)}{9 (w-1) (w+1)}
\nonumber\\&\quad
\times\Big[(166-4 n_l) G(-1,-1,0,w)
+(94-4 n_l) G(-1,0,-1,w)+(110-4 n_l) G(-1,0,0,w)
\nonumber\\&\quad
+(2 n_l-63) G(-1,0,1,w)
+(4 n_l-34) G(-1,1,0,w)+(8 n_l-260) G(0,-1,-1,w)
\nonumber\\&\quad
+(4 n_l-110) G(0,0,-1,w)
+(2 n_l-55) G(0,0,1,w)+(45-2 n_l) G(0,1,1,w)
\nonumber\\&\quad
+(4 n_l-34) G(1,-1,0,w)
+(110-4 n_l) G(1,0,-1,w)+(55-2 n_l) G(1,0,0,w)
\nonumber\\&\quad
+(4 n_l-110) G(1,1,0,w)
-20 G(0,-1,1,w)-20 G(0,1,-1,w)\Big]
\nonumber\\&\quad
+\frac{31 w^4+86 w^3-49 w^2+86 w+31}{9 (w-1) (w+1)^3}\Big[6 G(x_1,0,-1,w)-4 G(x_1,0,0,w)+3 G(x_1,0,1,w)
\nonumber\\&\quad
-G(x_1,1,0,w)+6 G(x_2,0,-1,w)-4 G(x_2,0,0,w)+3 G(x_2,0,1,w)-G(x_2,1,0,w)\Big]
\nonumber\\&\quad
+\frac{4 n_l \left(w^5+w^4-w-1\right)-118 w^5+480 w^4-459 w^3+331 w^2+628 w+162}{9 (w-1)^2 (w+1)^3}
\nonumber\\&\quad
\times\Big[2 G(0,0,-1,w)-G(0,0,1,w)\Big]+\frac{1}{18 (w-1)^2 (w+1)^3}
\nonumber\\&\quad
\times\Big[2 (w+1) \left(20 n_l \left(w^4-1\right)-383 w^4+1586 w^3-2786 w^2+1586 w+1149\right) 
\nonumber\\&\quad
\times G(-1,0,0,w)
-32 (w+1) \left(n_l \left(w^4-1\right)+3 w^4+32 w^3-123 w^2+32 w+40\right)
\nonumber\\&\quad
\times G(0,-1,0,w)+8\left(-4 n_l (w+1)^2 \left(w^2+1\right)+63 w^4+120 w^3+130 w^2+120 w+63\right) 
\nonumber\\&\quad
\times 
(w-1)G(1,0,-1,w)
\nonumber\\&\quad
-\left(-24 n_l (w+1)^2 \left(w^2+1\right)+183 w^4-1220 w^3+1662 w^2+4444 w+1459\right)
\nonumber\\&\quad
\times(w-1) G(1,0,0,w)
\nonumber\\&\quad
+2\left(-8 n_l (w+1)^2 \left(w^2+1\right)+126 w^4+240 w^3+260 w^2+240 w+126\right) 
\nonumber\\&\quad
\times(w-1) G(1,0,1,w)
\nonumber\\&\quad
-8 \left(4 n_l \left(w^5+w^4-w-1\right)-118 w^5+480 w^4-459 w^3+331 w^2+628 w+162\right) 
\nonumber\\&\quad
\times G(0,0,-1,w)
\nonumber\\&\quad
-2 \left(12 n_l \left(w^5+w^4-w-1\right)+89 w^5+319 w^4-1018 w^3-1278 w^2+760 w+488\right) 
\nonumber\\&\quad
\times G(0,1,0,w)
+2 (w-1) \left(139 w^4+284 w^3+274 w^2+284 w+139\right) G(1,1,0,w)
\nonumber\\&\quad
+\left(-11 w^5-43 w^4-203 w^3-157 w^2+26 w+4\right) G(0,0,0,w)\Big]
\nonumber\\&\quad
+\frac{8 i \pi  \left(w^2+1\right)}{9 \left(w^2-1\right)}\Big[(2 n_l-83) G(-1,-1,w)-(2 n_l-17) G(-1,1,w)
\nonumber\\&\quad
-(2 n_l-17) G(1,-1,w)\Big]
+\frac{i \pi  \left(31 w^4+86 w^3-49 w^2+86 w+31\right)}{9 (w-1) (w+1)^3}\Big[4 G(x_1,0,w)
\nonumber\\&\quad
+G(x_1,1,w)+4 G(x_2,0,w)
+G(x_2,1,w)\Big]
\nonumber\\&\quad
+\frac{8 \pi ^2 \left(31 w^4+86 w^3-49 w^2+86 w+31\right)}{27 (w-1) (w+1)^3}\Big[G(x_1,w)+G(x_2,w)\Big]
\nonumber\\&\quad
-\frac{i \pi ^3 \left(22 w^5+66 w^4+205 w^3+131 w^2-37 w-3\right)}{54 (w-1)^2 (w+1)^3}+\frac{2\left(w^2-4 w+1\right) \pi ^2 \log (2)}{9 (w+1)^2}
\nonumber\\&\quad
-\frac{\left(6 n_l \left(w^5+w^4-w-1\right)-159 w^5+317 w^4+44 w^3+844 w^2+510 w+172\right)\zeta (3)}{9 (w-1)^2 (w+1)^3}
\nonumber\\&\quad
+\frac{\pi ^2}{54 (w-1)^2 (w+1)^3}\Big[-(w+1)(32 n_l \left(w^4-1\right)-217 w^4+6354 w^3-10770 w^2
\nonumber\\&\quad
+6354 w+3271) G(-1,w)+\left(33 w^5+109 w^4+408 w^3+288 w^2-63 w-7\right) G(0,w)
\nonumber\\&\quad
-\left(24 n_l \left(w^5+w^4-w-1\right)+311 w^5+3473 w^4-6172 w^3-5972 w^2+5317w+2275\right) 
\nonumber\\&\quad
\times G(1,w)\Big]
\nonumber\\&\quad
+\frac{i \pi }{18 (w-1)^2 (w+1)^3}\Big[-2 (w+1) 
\nonumber\\&\quad
\times(4 n_l \left(w^4-1\right)+57 w^4+1586 w^3-2786 w^2+1586 w+709) G(-1,0,w)
\nonumber\\&\quad
+32 (w+1) \left(n_l \left(w^4-1\right)+3 w^4+32 w^3-123 w^2+32 w+40\right) G(0,-1,w)
\nonumber\\&\quad
+\left(11 w^5+43 w^4+203 w^3+157 w^2-26 w-4\right) G(0,0,w)
\nonumber\\&\quad
+2 \left(12 n_l \left(w^5+w^4-w-1\right)+89 w^5+319 w^4-1018 w^3-1278 w^2+760 w+488\right) 
\nonumber\\&\quad
\times G(0,1,w)
\nonumber\\&\quad
-\left(8 n_l \left(w^5+w^4-w-1\right)+257 w^5+1843 w^4-2882 w^3-2782 w^2+2545 w+1019\right) 
\nonumber\\&\quad
\times G(1,0,w)
\nonumber\\&\quad
+\left(-32 n_l \left(w^5+w^4-w-1\right)+602 w^5+590 w^4+20 w^3-20 w^2-590 w-602\right) 
\nonumber\\&\quad
\times G(1,1,w)
\Big]
-\frac{2n_l-55}{12} \log^2\left(\frac{\mu ^2}{m_H^2}\right) +\frac{2n_l-55}{9}\log\left(\frac{\mu^2}{m_H^2}\right)\Big(4 G(-1,w)-G(1,w)\Big)
\nonumber\\&\quad
+\frac{i (2n_l-55) w \left(3 w^3-w^2-8 w-7\right)}{9 (w-1) (w+1)^3}\log\left(\frac{\mu^2}{m_H^2}\right)\Big(\pi +i G(0,w)\Big)
\nonumber\\&\quad
-\frac{2}{9}\Big[16 n_l G(-1,-1,w)+2n_lG(1,1,w)-392 G(-1,-1,w)+24 G(-1,1,w)
\nonumber\\&\quad
+24 G(1,-1,w)-43 G(1,1,w)\Big]-\frac{2 w}{3 \left(w^2-1\right)}\Big[-2 G(0,0,w)+2 i \pi  G(0,w)\Big]
\nonumber\\&\quad
+\frac{8 \left(n_l \left(11 w^2-12 w-1\right)-321 w^2+471 w-45\right)}{27 \left(w^2-1\right)}G(-1,0,w)
\nonumber\\&\quad
+\frac{1}{27 (w-1) \left(w^3+1\right)^3}\Big[\Big(
4 n_l \left(w^2-w+1\right)^3 \left(11 w^4+4 w^3-20 w^2-44 w-13\right)
\nonumber\\&\quad
-6 (41 w^{10}-626 w^9+957 w^8-2300 w^7+2753 w^6-4566 w^5+3857 w^4-3404 w^3
\nonumber\\&\quad
+957 w^2-258 w-327)\Big)G(0,-1,w)-\Big(2 n_l (w+1)^2 \left(w^2-w+1\right)^3 \left(7 w^2-24 w+13\right)
\nonumber\\&\quad
-3 (289 w^{10}-450 w^9+449 w^8+1224 w^7-2925 w^6+4408 w^5-3039 w^4+1338 w^3
\nonumber\\&\quad
+449 w^2-488 w+327)\Big)G(0,1,w) - \Big(4 n_l \left(w^2-w+1\right)^3 \left(w^4-w^3-28 w^2+5 w+4\right)
\nonumber\\&\quad
-3 (57 w^{10}-245 w^9-516 w^8+2349 w^7-5403 w^6+6486 w^5-5517 w^4+2463 w^3
\nonumber\\&\quad
-516 w^2
-283 w+95)\Big)G(1,0,w)\Big]
+\frac{1}{54 (w-1)^2 (w+1)^4 \left(w^2-w+1\right)^3}
\nonumber\\&\quad
\times\Big[
-2 (6 n_l+527) w^{12}+(336 n_l-12575) w^{11}+(32735-936 n_l) w^{10}
\nonumber\\&\quad
+24 (57 n_l-2240) w^9
+(31063-756 n_l) w^8+(12305-684 n_l) w^7
\nonumber\\&\quad
+4 (423 n_l-12790) w^6+(44291-1440 n_l) w^5
+(360 n_l-15539) w^4
\nonumber\\&\quad
+24 (13 n_l-174) w^3+(6221-348 n_l) w^2+(108 n_l-1061) w+38
\Big]G(0,0,w)
\nonumber\\&\quad
+\frac{\pi ^2}{54 (w-1)^2 (w+1)^4 \left(w^2-w+1\right)^3}\Big[n_l \left(w^2-w+1\right)^3 
(23 w^6-162 w^5-71 w^4+164 w^3
\nonumber\\&\quad
+45 w^2-2 w+3)+281 w^{12}+9318 w^{11}-21794 w^{10}
+34216 w^9-18084 w^8-7596 w^7
\nonumber\\&\quad
+31012 w^6-26382 w^5+11106 w^4-160 w^3-1732 w^2-84 w-117\Big]
\nonumber\\&\quad
-\frac{8 i \pi  \left(n_l \left(11 w^2-12 w-1\right)-321 w^2+471 w-45\right)}{27 \left(w^2-1\right)}G(-1,w)
\nonumber\\&\quad
+\frac{i \pi }{54 (w-1)^2 (w+1)^4 \left(w^2-w+1\right)^3}\Big[
2 (6 n_l+527) w^{12}+(12575-336 n_l) w^{11}
\nonumber\\&\quad
+(936 n_l-32735) w^{10}-24 (57 n_l-2240) w^9+(756 n_l-31063) w^8+(684 n_l-12305) w^7
\nonumber\\&\quad
+(51160-1692 n_l) w^6+(1440 n_l-44291) w^5+(15539-360 n_l) w^4+(4176-312 n_l) w^3
\nonumber\\&\quad
+(348 n_l-6221) w^2+(1061-108 n_l) w-38
\Big]G(0, w)
\nonumber\\&\quad
+\frac{i \pi }{27 (w-1) \left(w^3+1\right)^3}\Big[4 n_l \left(w^2-w+1\right)^3 \left(w^4-w^3-28 w^2+5 w+4\right)
\nonumber\\&\quad
-3 (57 w^{10}-245 w^9-516 w^8+2349 w^7-5403 w^6+6486 w^5-5517 w^4
\nonumber\\&\quad
+2463 w^3-516 w^2-283 w+95)\Big]G(1, w)
\nonumber\\&\quad
-\frac{\left(n_l\left(22 w^2+50 w+22\right)-614 w^2-1393 w-614\right)}{18 (w+1)^2}\log \left(\frac{\mu ^2}{m_H^2}\right)
\nonumber\\&\quad
+\frac{1}{27 \left(w^3+1\right)^2}\Big[
2 n_l \left(w^2-w+1\right)^2 \left(59 w^2+136 w+59\right)
\nonumber\\&\quad
-3 \left(993 w^6+736 w^5-1424 w^4+4130 w^3-1424 w^2+736 w+993\right)
\Big]G(-1,w)
\nonumber\\&\quad
+\frac{1}{324 (w-1) (w+1)^3 \left(w^2-w+1\right)^2}\Big[
(31015-1332 n_l) w^8+6 (170 n_l-2271) w^7
\nonumber\\&\quad
+(1440 n_l-81929) w^6+(192374-5112 n_l) w^5+6 (916 n_l-34731) w^4
\nonumber\\&\quad
-22 (72 n_l-2963) w^3
+(14605-756 n_l) w^2+6 (242 n_l-7905) w-503\Big]G(0,w)
\nonumber\\&\quad
+\frac{1}{54 (w+1)^2 \left(w^2-w+1\right)^2}\Big[-14n_l \left(w^3+1\right)^2
\nonumber\\&\quad
+3 \left(235 w^6-406 w^5+764 w^4-684 w^3+764 w^2-406 w+235\right)\Big]G(1,w) 
\nonumber\\&\quad
+ \frac{i \pi }{324 (w-1) (w+1)^3 \left(w^2-w+1\right)^2}\Big[
(1332 n_l-31015) w^8
\nonumber\\&\quad
-6 (170 n_l-2271) w^7+(81929-1440 n_l) w^6+2 (2556 n_l-96187) w^5
\nonumber\\&\quad
-6 (916 n_l-34731) w^4
+22 (72 n_l-2963) w^3+(756 n_l-14605) w^2
\nonumber\\&\quad
-6 (242 n_l-7905) w+503
\Big]
\nonumber\\&\quad
+\frac{-18 n_l \left(195 w^2+374 w+195\right)+98681 w^2+204550 w+98681}{1296 (w+1)^2}
\end{align}
with
\begin{align}
x_1 = l^{\text{NP1}}_{4} = \frac{1+\sqrt{3}i}{2},\quad x_2 = l^{\text{NP1}}_{5} = \frac{1-\sqrt{3}i}{2}\,.
\end{align}
Here the color factors $C_F=4/3$, $C_A=3$, and $T_R=1/2$ have been substituted by their numeric values,
and $n_l$ denotes the number of light quark flavors.

The result of $X_{2,b\bar{b}b\bar{b}}^{y_by_b}$ is given by 
\begin{align}
X_{2,b\bar{b}b\bar{b}}^{y_by_b} &= \frac{8}{3 \pi \beta  (z-4) z}\Big[\frac{F^{4b}_1(z) \left(-207 z^2-3964 z-11264\right)}{55296}
\nonumber\\&\quad
+\frac{F^{4b}_2(z) \left(479 z^2+18860 z-34528\right)}{82944}
-\frac{F^{4b}_3(z) (z-4) (2491 z+3012)}{20736}
\nonumber\\&\quad
+\frac{F^{4b}_4(z) \left(-55 z^3+254 z^2+2328 z-9216\right)}{3456 \beta  z}
+\frac{F^{4b}_5(z) \left(-5 z^2+24 z-32\right)}{2304}
\nonumber\\&\quad
+\frac{\beta  F^{4b}_6(z) \left(-5 z^2+10 z+384\right)}{1152}
+\frac{F^{4b}_7(z) \left(5 z^2-42 z+90\right)}{576}
\nonumber\\&\quad
+\frac{\beta  F^{4b}_8(z) \left(17 z^2-58 z-384\right)}{1152}+\frac{F^{4b}_9(z) \left(z^3-6 z^2+z-192\right)}{288 z}
\nonumber\\&\quad
+\frac{F^{4b}_{10}(z) \left(z^2-3 z-16\right)}{144}
-\frac{\beta  F^{4b}_{12}(z) (z-2)^2}{144}\Big]\,,
\label{eq:Hto4b}
\end{align}
where $\beta = \sqrt{1-4/z}$ measures the velocity of the bottom quark. The master integrals $F^{4b}_1(z)$, $F^{4b}_2(z)$ and $F^{4b}_3(z)$ are linear combinations of complete elliptic integrals or their derivatives, and the other $F^{4b}_i(z)$ can be expressed by one-fold integrals of
MPLs and complete elliptic integrals. The complete expressions of all $F^{4b}_i(z)$ can be found in \cite{Lee:2019wwn}. We also provide these expressions in electronic form in the ancillary file attached to this paper. %$X_{2,b\bar{b}b\bar{b}}^{y_by_b}$ will vanish if $z\rightarrow4\,(m_H\rightarrow4m_b)$. 

\subsection{Asymptotic expansion}

It is interesting to explore the decay width in various limits.
First, we study the small bottom quark mass limit, i.e., $m_b^2\rightarrow 0$ or $z\rightarrow\infty$,
while the Yukawa coupling is still kept finite.
The NLO coefficient becomes 
\begin{align}
X_1^{y_by_b}|_{z\rightarrow \infty} = 2\log\left(\frac{\mu ^2}{m_H^2}\right)+\frac{17}{3}+\frac{12\log(z)+10}{z}
+\mathcal{O}(z^{-2})\,.
\end{align}
Therefore, it is safe to take the continuous limit of $m_b=0$.
This property is in accordance with the Kinoshita-Lee-Nauenberg theorem \cite{Kinoshita:1962ur,Lee:1964is}.
Here the parameter $m_b$ can be considered as a regulator for infrared divergences.
Though the virtual and real corrections contain 
$\log(z)$ terms induced by collinear gluons separately,
the dependence on the regulator cancels in the decay width at $\mathcal{O}(z^{0})$.
Notice that there are large $\log(z)$ terms at $\mathcal{O}(z^{0})$ if the Yukawa coupling is renormalized in the on-shell scheme.

In the limit of $z\rightarrow\infty$, the NNLO coefficient $\tilde{X}_{2,b\bar{b}}^{y_by_b}$ reads
\begin{align}
&\tilde{X}_{2,b\bar{b}}^{y_by_b}|_{z\rightarrow \infty} = -\frac{\log^3 \left(z\right)}{27} +\frac{19\log ^2\left(z\right)}{54}+\frac{2\zeta (3)\log \left(z\right)}{9}+\frac{\pi ^2 \log \left(z\right)}{54}
\nonumber\\&\quad
-\frac{503\log \left(z\right)}{324}+\frac{55-2n_l}{12} \log ^2\left(\frac{\mu ^2}{m_H^2}\right)+\frac{307-11n_l}{9} \log \left(\frac{\mu ^2}{m_H^2}\right)+\frac{2(3n_l-86)\zeta (3)}{9}
\nonumber\\&\quad
+\frac{(3n_l-98) \pi ^2}{54}-\frac{19 \pi ^4}{1620}+\frac{2\log (2)\pi ^2 }{9}+\frac{98681-3510n_l}{1296}
\nonumber\\&\quad
+\frac{1}{z}\bigg(-\frac{5 \log ^4(z)}{36}+\frac{\log ^3(z)}{18}+\frac{\pi ^2 \log ^2(z)}{18}-\frac{(2n_l-9)\log ^2(z)}{2} -\frac{7 \pi ^2\log (z)}{54}
\nonumber\\&\quad
-(2n_l-55) \log (z) \log\left(\frac{\mu ^2}{m_H^2}\right)-\frac{(117 n_l-3625) \log (z)}{27}-4 \zeta (3) \log (z)
\nonumber\\&\quad
-\frac{71 \pi ^4}{540}+\frac{4\pi ^2 \log (2)}{3}-\frac{5(2n_l-55)}{6}\log\left(\frac{\mu ^2}{m_H^2}\right)
+\frac{430 \zeta (3)}{9}-\frac{(9n_l-175)\pi ^2}{9}
\nonumber\\&\quad
+\frac{36829-1548n_l}{324}\bigg)+\mathcal{O}(z^{-2}),
\label{eq:asy2b}
\end{align}
where we have shown explicitly the first two orders in the expansion.
It is obvious that there are large logarithms $\log^i (z) $ with $i\le 3$ at $\mathcal{O}(z^0)$.
This means that the contribution of this part with two bottom quarks in the final state to the decay width is divergent in the massless limit.

\begin{figure}[ht]
	\centering
	\begin{minipage}{0.45\linewidth}
		\centering
		\includegraphics[width=0.85\linewidth]{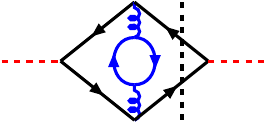}
        \caption*{(a)}
	\end{minipage}
	\begin{minipage}{0.45\linewidth}
		\centering
		\includegraphics[width=0.85\linewidth]{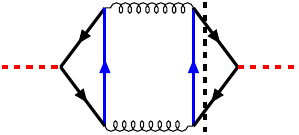}
        \caption*{(b)}
	\end{minipage}
	\begin{minipage}{0.45\linewidth}
	       \centering
	       \includegraphics[width=0.85\linewidth]{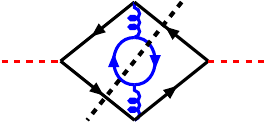}
               \caption*{(c)}
	\end{minipage}
 	\begin{minipage}{0.45\linewidth}
	       \centering
	       \includegraphics[width=0.85\linewidth]{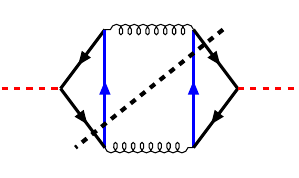}
        \caption*{(d)}
	\end{minipage}
\caption{Typical cut diagrams that have large logarithms $\log^i(z)$. The thick black and red lines stand for the massive bottom quark and the Higgs boson, respectively. And the blue line represents the particle that is soft. }
\label{soft_ThreeLoop}
\end{figure}

These logarithms $\log^i (z) $ arise from the diagrams with a bottom quark loop, e.g., see the diagram (a) in figure \ref{soft_ThreeLoop}.
These diagrams contain up to $1/\epsilon^3$ infrared divergences in the massless case, 
which are converted to $\log^i (z) , i\le 3$, in the massive bottom quark case.

It is interesting to note the $\log^4 (z)$ term at order $\mathcal{O}(z^{-1})$.
It appears due to a different reason.
It comes from the contribution of diagrams with two soft quarks exchanged between two separate directions,
e.g., see the diagram (b) in figure \ref{soft_ThreeLoop}. 

The other NNLO coefficient $X_{2,b\bar{b}b\bar{b}}^{y_by_b}$ consists of master integrals that depend on elliptic integrals.
We present the asymptotic expansion results of the master integrals in appendix \ref{sec:asyMI}.
Based on these results, we obtain the asymptotic expansion,
\begin{align}
&X_{2,b\bar{b}b\bar{b}}^{y_by_b}|_{z\rightarrow \infty} = \frac{\log^3(z)}{27}-\frac{19\log^2(z)}{54}-\frac{2\zeta (3)\log(z)}{9} 
-\frac{\pi ^2\log(z)}{54}+\frac{503\log(z)}{324}
\nonumber\\&\quad
+\frac{19 \pi ^4}{1620}-\frac{2\log(2)\pi^2}{9} +\frac{5 \zeta (3)}{18}+\frac{31 \pi ^2}{108}-\frac{2491}{648}
\nonumber\\&\quad
+\frac{1}{z}\bigg(\frac{\log^4(z)}{18}-\frac{\log ^3(z)}{18}-\frac{2\pi ^2\log ^2(z)}{9}+\frac{7\pi ^2 \log (z)}{54}-\frac{\log ^2(z)}{3}
+4 \zeta (3) \log (z)
\nonumber\\&\quad
+\frac{211 \log (z)}{54}+\frac{13 \pi ^4}{270}-\frac{7 \zeta (3)}{9}+\frac{19 \pi ^2}{18}-\frac{8353}{324}
\bigg)
+\mathcal{O}(z^{-2})\,.
\label{eq:asy4b}
\end{align}
It can be seen that  $X_{2,b\bar{b}b\bar{b}}^{y_by_b}$ also have large logarithms when $z\rightarrow\infty$. 
The logarithms $\log^i (z),i\le 3,$ at $\mathcal{O}(z^0)$ are obtained by calculating the diagrams with a soft gluon splitting to a soft bottom quark pair; see the diagram (c) in figure \ref{soft_ThreeLoop}.
In contrast, the logarithm $\log^4 (z)$ at $\mathcal{O}(z^{-1})$ arises from the diagrams with a collinear quark splitting to a soft quark and a collinear gluon 
which splits into a collinear quark and a soft quark further, such as the diagram (d) in figure \ref{soft_ThreeLoop}.

Combining the results in eq.(\ref{eq:asy2b}) and eq.(\ref{eq:asy4b}),  we obtain
\begin{align}
&X_2^{y_by_b}|_{z\rightarrow\infty\,(m_b^2\rightarrow0)} = \frac{55-2n_l}{12} \log ^2\left(\frac{\mu ^2}{m_H^2}\right)+\frac{307-11n_l}{9} \log \left(\frac{\mu ^2}{m_H^2}\right)+\frac{(4n_l-113) \zeta (3)}{6}
\nonumber\\&\quad
+\frac{(2n_l-55) \pi ^2}{36}+\frac{10411-390n_l}{144}+\frac{1}{z}\bigg(
-\frac{\log^4(z)}{12}-\frac{\pi ^2 \log ^2(z)}{6}-\frac{ (6n_l-25) \log ^2(z)}{6}
\nonumber\\&\quad
-\frac{(26n_l-829 )\log (z)}{6}
-(2n_l-55) \log (z) \log\left(\frac{\mu ^2}{m_H^2}\right)-\frac{5(2n_l-55)}{6}\log\left(\frac{\mu ^2}{m_H^2}\right)
-\frac{\pi ^4}{12}
\nonumber\\&\quad
+\frac{4\pi ^2 \log (2)}{3}+47 \zeta (3)-\frac{ (2n_l-41)\pi ^2}{2}+\frac{791-43n_l}{9}
\bigg)
+\mathcal{O}(z^{-2})\,.
\end{align}
The large logarithms of $z$ at $\mathcal{O}(z^0)$ cancel, as expected.
However, the large logarithms at $\mathcal{O}(z^{-1})$ survive. 
The structure of this kind of large logarithm at higher orders in $\als$ is not well understood.
It is promising to explore this issue using the effective field theory, which we leave for future study.
After adopting the same renormalization scheme, the above expression is consistent with the result in ref.~\cite{Harlander:1997xa}, which was obtained with the method of large momentum expansion.

It is also interesting to investigate the threshold limit of bottom quark pair production, where the coefficients $X_1^{y_by_b}$ and $X_2^{y_by_b}$ are expanded in the limit of $ m_H\rightarrow 2m_b$ or $\beta\rightarrow 0$. 
Expanding the full analytic results, we have
\begin{align}
X_1^{y_by_b}|_{\beta\rightarrow 0} = \frac{2 \pi ^2}{3 \beta } +
4\log(2) +\frac{4}{3}+2\log\left(\frac{\mu ^2}{m_H^2}\right)+\frac{2 \pi ^2 \beta }{3}+\mathcal{O}\left(\beta^2\right),
\end{align}
and 
\begin{align}
&X_2^{y_by_b}|_{\beta\rightarrow0 } = \frac{1}{\beta^2}\left(\frac{4 \pi ^4}{27}+\frac{4 \pi ^2}{9}\right)
+\frac{1}{\beta}\bigg(
\frac{\left(339-22 n_l\right)\pi ^2}{54}+\frac{22\log(2)\pi^2}{9}
\nonumber\\ &\quad
+\frac{\left(2 n_l-33\right)\pi^2\log(\beta)}{9}+\frac{\left(55-2 n_l\right)\pi ^2}{18}\log\left(\frac{\mu^2}{m_H^2}\right)
\bigg)
\nonumber\\ &\quad
+\bigg(\frac{8 \pi ^4}{27}+\frac{(139-3 n_l)\pi^2}{27}-\frac{100\log (\beta )\pi^2}{27}-\frac{130\log (2)\pi^2}{27}
-\frac{181 \zeta (3)}{9}
\nonumber\\ &\quad
+\frac{(55-2 n_l)}{12}\log ^2\left(\frac{4\mu ^2}{m_H^2}\right)+\frac{(57-2 n_l)}{4}\log\left(\frac{\mu ^2}{m_H^2}\right)+\frac{(163-6 n_l)\log (2)}{6}
\nonumber\\ &\quad
+\frac{(13993-506 n_l)}{432}
\bigg)
+\beta\bigg(\frac{\left(6 n_l-303\right)\pi ^2}{162} +\frac{\left(6 n_l+101\right)\pi ^2\log (\beta )}{27}
\nonumber\\  &\quad
+\frac{\left(55-2 n_l\right) \pi ^2}{18}\log\left(\frac{\mu ^2}{m_H^2}\right)
+\frac{418\log (2)\pi ^2 }{27}
\bigg)+\mathcal{O}\left(\beta^2\right).
\end{align}
In obtaining the above compact expressions, we have extensively used the relations among MPLs 
%with arguments of $\{0,\pm1,\pm1\pm\sqrt{3}i\}$ 
in \cite{Henn:2015sem,Frellesvig:2016ske}. We check that the coefficients of $1/\beta^{2}$ and $1/\beta$ are 
consistent with the results of two-loop form factors \cite{Ablinger:2017hst},
which means that the power divergences at $\beta\to 0$ in $X_1^{y_by_b}$ and $X_2^{y_by_b}$ come only from virtual corrections.
Actually, these velocity-enhanced terms are due to the Coulomb potential interaction between the bottom quark pair.
They violate the convergence of perturbative expansion \footnote{The perturbative expansion has a structure $\beta^3 \sum_{n=0} c_n \alpha_s^{n}/\beta^{n} $ in the limit of $\beta\to 0$. It becomes divergent for $n>3$.} and need to be resummed to all orders in $\alpha_s$ by calculating the corresponding Coulomb Green function \cite{Beneke:1999qg}.
Notice that the coefficients of $\pi^2/\beta$ in $X_1^{y_by_b}$ and $\pi^4/\beta^2$ in  $X_2^{y_by_b}$ agree with those in $e^+ e^- \to t\bar{t}$ production near threshold \cite{Czarnecki:1997vz},
but the $\pi^2/\beta^2$ term is different.
They can be reproduced by calculating the imaginary part of the first a few terms of the zero-distance Coulomb Green function,
\begin{align}
{\rm Im}& ~G_0 (E)= 
-{\rm Im} \int\frac{d^{D-1} \bff{p}_1}{(2\pi)^{D-1}}
\frac{\bff{p}_1\cdot \bff{p}_1}{E-\frac{\bff{p}_1^2}{m_b}}\,
+{\rm Im} \int\frac{d^{D-1} \bff{p}_1}{(2\pi)^{D-1}}
\frac{d^{D-1} \bff{p}_2}{(2\pi)^{D-1}}
\frac{\bff{p}_1\cdot \bff{p}_2}{E-\frac{\bff{p}_1^2}{m_b}}\,
\frac{g_s^2 C_F}{(\bff{p}_1-\bff{p}_2)^2} \,
\frac{1}{E-\frac{\bff{p}_2^{2}}{m_b}}
\nn\\
&-{\rm Im} \int\frac{d^{3} \bff{p}_1}{(2\pi)^{3}}
\frac{d^{3} \bff{p}_2}{(2\pi)^{3}}
\frac{d^{3} \bff{p}_3}{(2\pi)^{3}}
\frac{\bff{p}_1\cdot \bff{p}_3}{E-\frac{\bff{p}_1^2}{m_b}}\,
\frac{g_s^2 C_F}{(\bff{p}_1-\bff{p}_2)^2} \,
\frac{1}{E-\frac{\bff{p}_2^{2}}{m_b}}
\frac{g_s^2 C_F}{(\bff{p}_2-\bff{p}_3)^2} \,
\frac{1}{E-\frac{\bff{p}_3^{2}}{m_b}}+\mathcal{O}(g_s^6)
\nn\\
&= \frac{\beta^3 m_b^4}{4\pi}\left(1+\frac{\als C_F}{\pi}\frac{\pi^2}{2\beta}+\left(\frac{\als C_F}{\pi}\right)^2\left(\frac{\pi^4}{12\beta^2}+\frac{\pi^2}{4\beta^2}\right)+\mathcal{O}(\als^3)\right)
\end{align}
with $E=m_b\beta^2+\mathcal{O}(\beta^4)$.
The $\bff{p}_i\cdot \bff{p}_j$ numerator arises because the scalar current $\bar{\psi}_b \psi_b $ with $\psi_b$ the bottom quark field matches to the current
$j^{(s)}=\psi^{\dagger} \bm{\sigma}\cdot i\bff{D} \chi /m_b $ in non-relativistic QCD where $\psi$ and $\chi$ represent the two-component quark and anti-quark field, respectively.
The all-order result of this Coulomb Green function can be found in refs. \cite{Solovtsov:2005xi,Beneke:2013kia}.

\section{Numerical results}
\label{sec:numerical}

We adopt the SM input parameters \cite{ParticleDataGroup:2022pth}
\begin{align}
    \overline{m_b}\,(\overline{m_b}) &= 4.18 ~{\rm GeV}, & m_b & = 5.07~{\rm GeV},  \nn \\
    m_H & = 125.09 ~{\rm GeV}, & m_Z & = 91.1876 ~{\rm GeV}, \\
    \als(m_Z) & = 0.1181, & G_F & = 1.166378 \times 10^{-5}~{\rm GeV}^{-2}, \nn
    \label{eq:input}    
\end{align}
where the on-shell value of $m_b$ is obtained from $\overline{m_b}(\overline{m_b})$ in the $\overline{\rm MS}$ scheme with the four-loop mass relation.
The package {\tt RunDec} \cite{Chetyrkin:2000yt,Herren:2017osy} was used to obtain $\overline{m_b}$ at other scales, e.g., 
$\overline{m_b}\,(m_H/2) = 2.956$ GeV, $\overline{m_b}\,(m_H) = 2.784$ GeV and $\overline{m_b}\,(2m_H) = 2.639$ GeV.

\begin{table}[tb]
	\centering
	\scalebox{1.0}{
		\begin{tabular}{llll}
			\toprule
			$\quad\quad\mu$&
			$\quad\frac{1}{2}m_H$ &
			$\quad m_H$ &
			$\quad 2m_H$ \\
			\midrule
			$X_1^{y_by_b}$(our res.)&
			$+3.037513882$ &
			$+5.810102605$ &
			$+8.582691327$ \\
			\midrule
			$X_1^{y_by_b}$(Ref. \cite{Harlander:1997xa})&
			$+3.037513882$ &
			$+5.810102605$ &
			$+8.582691327$ \\
			\midrule
			$\tilde{X}_{2,b\bar{b}}^{y_by_b}$(our res.)&
			$-4.537634223$ &
			$+29.22467780$ &
			$+78.04118427$ \\
			\midrule
			${X}_2^{y_by_b}$(our res.)&
			$-3.181291730$ &
			$+30.58102030$ &
			$+79.39752676$ \\
			\midrule
                ${X}_2^{y_by_b}$(Ref. \cite{Harlander:1997xa})&
			$-3.181291730$ &
			$+30.58102030$ &
			$+79.39752676$ \\
			\bottomrule
		\end{tabular}
	}
	\caption{The numerical results for the $X_1^{y_by_b}$, $\tilde{X}_2^{y_by_b}$ and ${X}_2^{y_by_b}$ coefficients.} 
	\label{X1X2}
\end{table}

The numerical results of $X_1^{y_by_b}$ and ${X}_2^{y_by_b}$ at the scale $\mu=m_H/2,m_H, 2m_H$ are shown in table \ref{X1X2}. 
We also provide the results from the asymptotic expansion expression\footnote{The result was calculated with the Yukawa coupling in the on-shell scheme. We converted it to the result in the $\overline{\rm MS}$ scheme with the mass relation in \cite{Broadhurst:1991fy,Gray:1990yh,Marquard:2015qpa,Marquard:2016dcn}.} in \cite{Harlander:1997xa}, which have been calculated up to $\mathcal{O}(m_b^8/m_H^8)$.
Perfect agreements were found for both the NLO $X_1^{y_b y_b}$ and NNLO $X_2^{y_b y_b}$ coefficients.
The reason is that the expansion parameter $m_b^2/m_H^2 \approx 5\times 10^{-4} $ is rather small.
Actually, the difference between the  expansion up to $\mathcal{O}(m_b^2/m_H^2)$ and the exact result at the scale $\mu = m_H$ is already $0.002\%$  for ${X}_2^{y_by_b}$.

\begin{figure}[ht]
	\centering
	\begin{minipage}{0.47\linewidth}
		\centering
		\includegraphics[width=1.01\linewidth]{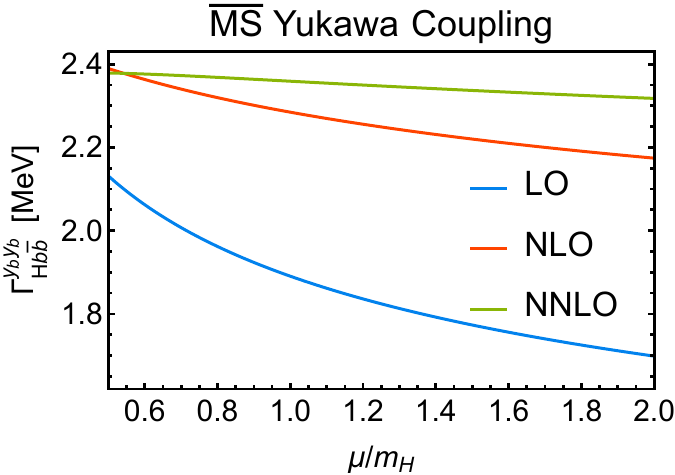}
	\end{minipage}
	\begin{minipage}{0.47\linewidth}
		\centering
		\includegraphics[width=1.0\linewidth]{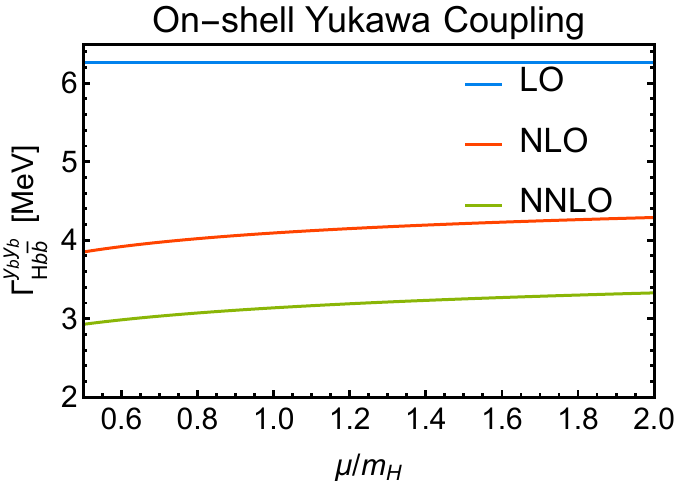}
	\end{minipage}
\caption{$\Gamma_{Hb\bar{b}}^{y_b y_b}$ at different values of $\mu$. The left (right) plot corresponds to the result with $\overline{\rm MS}$ (on-shell) Yukawa couplings. The LO, NLO and NNLO predictions are represented by the blue, red, and green lines, respectively.}
\label{fig:scl}
\end{figure}

Combining the corrections up to $\mathcal{O}(y_b^2\alpha_s^2)$, we obtain  $\Gamma^{y_by_b}_{Hb\bar{b}}$ up to NNLO. We show the results of $\Gamma^{y_by_b}_{Hb\bar{b}}$ with both the $\overline{\rm MS}$  and the on-shell Yukawa coupling in figure~\ref{fig:scl}.
The renormalization scale $\mu$ varies from $\mu = 1/2 m_H$ to $\mu = 2 m_H$.
In the $\overline{\rm MS}$ scheme,
the LO decay width exhibits a strong scale dependence, and it changes by about $23\%$ for $\mu$ in the range $ [1/2m_H, 2m_H]$.
The scale uncertainty is significantly reduced down to about $9\%$ and $3\%$ at NLO and  NNLO, respectively.
The NLO correction increases the LO decay width by a factor of $12\% - 28\%$ depending on the scale.  
The NNLO correction is very small at $\mu=m_H/2$ but can be as large as $7\%$ at $\mu=2m_H$.

The LO decay width in the on-shell scheme does not depend on the scale.
The scale uncertainty is about $11\%$ and $13\%$ at NLO and NNLO, respectively.
At a typical scale $\mu=m_H$, the NLO correction decreases the LO result by $35\%$, and the NNLO correction reduces the decay width further by $23\%$. 
This behavior indicates that the perturbative expansion with the on-shell Yukawa coupling converges slower than that with the $\overline{\rm MS}$ Yukawa coupling.

Comparing the two schemes,
one can find that the results in the on-shell scheme are larger than those in the $\overline{\rm MS}$ scheme.
The main reason is that the pole mass $m_b$ = 5.07 GeV is much larger than the $\overline{\rm MS}$ mass $\overline{m_b}(\mu=m_H)$ = 2.78 GeV.
However, higher-order corrections reduce the difference between the two schemes prominently,
signifying its importance in providing reliable theoretical predictions.

\begin{table}
    \centering
    \begin{tabular}{ccccccc}
        width & LO & NLO & 
        NNLO($y_b^2$) & NNLO($y_by_t$) & N$^3$LO($y_b^2$) & N$^4$LO($y_b^2$) \\
        \hline 
        $\overline{\rm MS}$ & 1.891$^{+0.241}_{-0.192}$ & 2.285$^{+0.105}_{-0.110}$ & 2.359$^{+0.020}_{-0.041}$ &  2.376$^{+0.026}_{-0.046}$& 2.379$^{+0.005}_{-0.015}$ &  2.377$^{+0.006}_{-0.006}$ \\
        \hline
        on-shell &  6.269&  4.092$^{+0.197}_{-0.242}$&  3.138$^{+0.191}_{-0.210}$&  3.193$^{+0.181}_{-0.197}$&  2.804$^{+0.112}_{-0.096}$& 2.649$^{+0.065}_{-0.049}$\\
    \end{tabular}
    \caption{Decay width of $\Gamma_{Hb\bar{b}}$ (in MeV) at different orders in QCD. The second row shows the results with the $\overline{\rm MS}$ Yukawa coupling while the third row shows the results with the on-shell Yukawa coupling. The renormalization scale uncertainties around $\mu=m_H$ are also shown.  The NNLO($y_by_t$), N$^3$LO($y_b^2$), and N$^4$LO($y_b^2$) results were calculated using the formulas in refs. \cite{Larin:1995sq,Chetyrkin:1996sr,Baikov:2005rw,Herzog:2017dtz}.} 
    \label{tab:totwidth}
\end{table}

Given that the scheme dependence constitutes the dominant theoretical uncertainty now, it is necessary to see how it changes as higher orders are taken into account.
We adopt the $\mathcal{O}(y_by_t\alpha_s^2)$ correction in the expansion of $m_b^2$/$m_H^2$ and $m_H^2$/$m_t^2$  \cite{Larin:1995sq}, 
and the $\mathcal{O}(y_b^2\alpha_s^3)$ and $\mathcal{O}(y_b^2\alpha_s^4)$ corrections in the massless bottom quark limit \cite{Chetyrkin:1996sr,Baikov:2005rw,Herzog:2017dtz}.
One can see that the central value for the decay width in the $\overline{\rm MS}$  scheme becomes stable quickly and that the scale uncertainty is decreased to $\pm 0.3\%$ after the N$^4$LO correction is included.
In contrast, the N$^3$LO correction in the on-shell scheme decreases the NNLO result by $12\%$,
and the N$^4$LO correction reduces the decay width by $5.5\%$ further.
The scale uncertainty is still $2\%$ at N$^4$LO.
All these features manifest that the results in the on-shell scheme have a bad convergence behavior. 
It is more appropriate to take the $\overline{\rm MS}$  scheme in this high-energy process.

\begin{figure}
    \centering
    \includegraphics[width=0.6\linewidth]{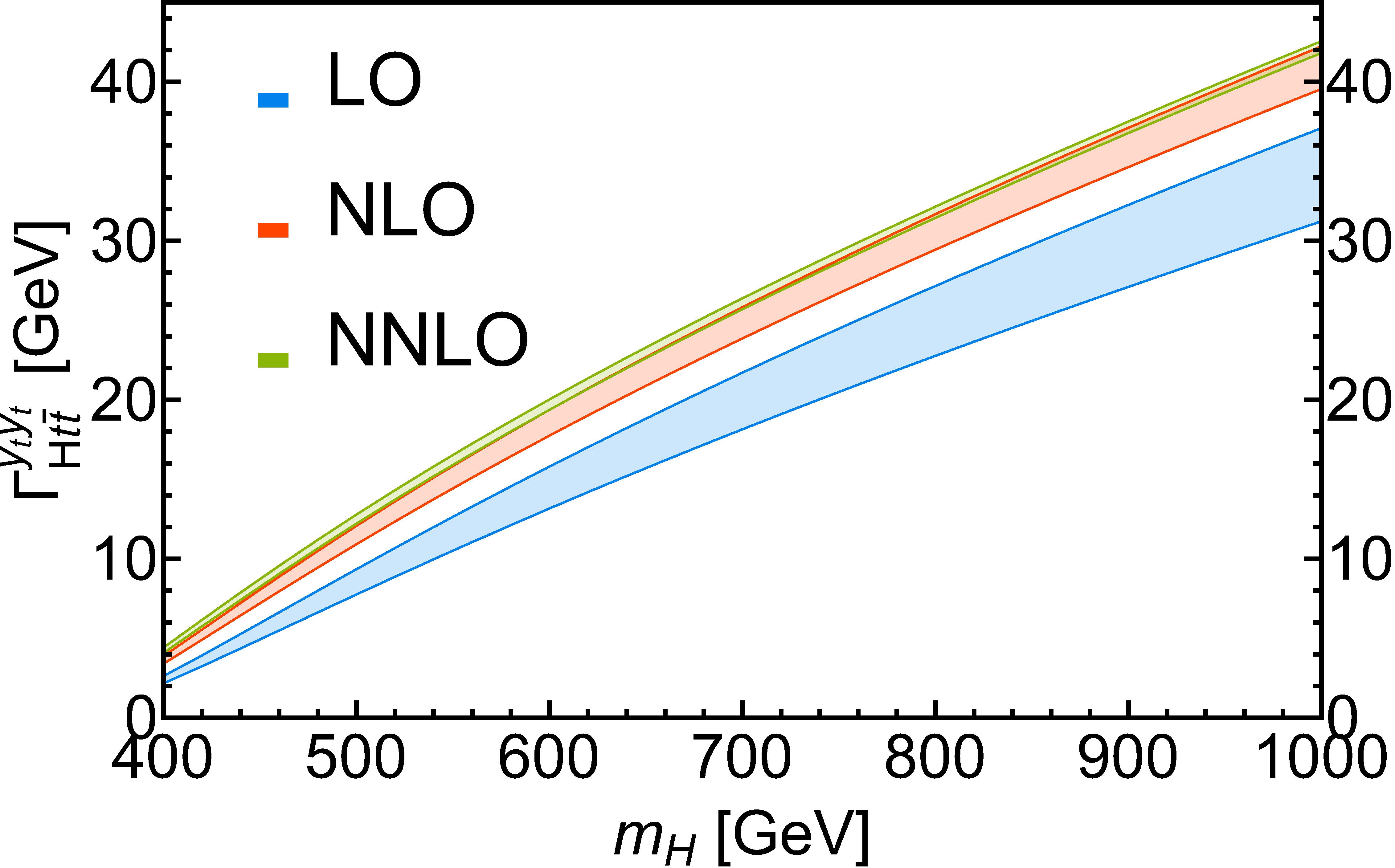}
    \caption{$\Gamma_{Ht\bar{t}}$ at different values of $m_H$. The LO, NLO and NNLO predictions with QCD scale uncertainties are represented by the blue, red, and green bands, respectively.}
    \label{fig:mH_mt}
\end{figure}

Our analytic results can also be applied to the decay of heavy scalar particles. If there is a new scalar Higgs with a mass of more than 350 GeV, it can decay to a top quark pair. 
In this process, we have five massless light quarks and one massive top quark. The decay width $\Gamma_{Ht\bar{t}}$ of $H\to t\bar{t}$  with $m_t = 172.69$ GeV and different values of $m_H$ are shown in figure~\ref{fig:mH_mt}.  
The NNLO correction increases the NLO result by $3\% - 17\%$ at $\mu = m_H$ as $m_H$ varies from 400 GeV to 1 TeV, and the scale uncertainty reduces from $7\% -13\%$ at NLO to $2\% - 8\%$ at NNLO.

\section{Conclusion}
\label{sec:conclusion}
We present an exact result of the decay width of $H\to b\bar{b}$ at $\mathcal{O}(y_b^2\alpha_s^2)$ with full dependence on $m_b$.
The result is obtained by calculating the imaginary part of the three-loop self-energy diagrams of the Higgs boson which have two bottom quarks in the propagators.
The master integrals are calculated using the method of differential equations combined with the choice of a canonical basis.
The result has been expressed in terms of multiple polylogarithms.
The contribution from the four bottom quarks is also calculated analytically and written as a linear combination of complete elliptic integrals and one-fold integrals of them.
The expansion of our analytical result in the small quark mass $m_b$ limit agrees with the previous computation in a different method.
The threshold expansion shows velocity enhancement which is caused by Coulomb potential interactions.
Though the structure is not the same as that in the top quark pair production near the threshold, we checked that the power divergent terms can be reproduced by calculating a new Coulomb Green function.
We discuss the numerical results in both the
$\overline{\rm MS}$ and on-shell renormalization schemes for the Yukawa coupling.
Higher-order QCD corrections reduce both the scale uncertainty and the difference between the two schemes.
The analytical results presented in this paper enable accurate and fast evaluation of the decay width of heavy scalar particles.

\section*{Acknowledgements}

We would like to thank Long Chen, Long-Bin Chen, Hai Tao Li, Xing Wang and Yu-Ming Wang for helpful discussions. We thank
Xin Guan for the communication on the implementation of packages {\tt AMFlow} and {\tt CalcLoop}. This work was supported in part by the National Science Foundation of China (grant Nos. 12005117, 12321005, 12375076) and the Taishan Scholar Foundation of Shandong province (tsqn201909011).
The topology diagrams in this paper were drawn using the TikZ-Feynman package \cite{Ellis:2016jkw}.

\appendix

\section{Topologies of the master integrals}
\label{appendix:topoMIs}

\begin{figure}[H]
	\centering
	\begin{minipage}{0.2\linewidth}
		\centering
		\includegraphics[width=1\linewidth]{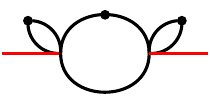}
		\caption*{$\M^{\text{NP1}}_{1}$}
	\end{minipage}
	\begin{minipage}{0.2\linewidth}
		\centering
		\includegraphics[width=1\linewidth]{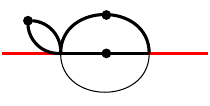}
		\caption*{$\M^{\text{NP1}}_{2}$}
	\end{minipage}
	\begin{minipage}{0.2\linewidth}
		\centering
		\includegraphics[width=1\linewidth]{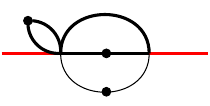}
		\caption*{$\M^{\text{NP1}}_{3}$}
	\end{minipage}
	\centering
	\begin{minipage}{0.2\linewidth}
		\centering
		\includegraphics[width=1\linewidth]{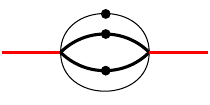}
		\caption*{$\M^{\text{NP1}}_{4}$}
	\end{minipage}
	\begin{minipage}{0.2\linewidth}
		\centering
		\includegraphics[width=1\linewidth]{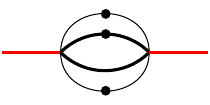}
		\caption*{$\M^{\text{NP1}}_{5}$}
	\end{minipage}
	\begin{minipage}{0.2\linewidth}
		\centering
		\includegraphics[width=1\linewidth]{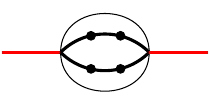}
		\caption*{$\M^{\text{NP1}}_{6}$}
	\end{minipage}
	\begin{minipage}{0.2\linewidth}
		\centering
		\includegraphics[width=1\linewidth]{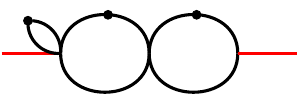}
		\caption*{$\M^{\text{NP1}}_{7}$}
	\end{minipage}
	\begin{minipage}{0.2\linewidth}
		\centering
		\includegraphics[width=1\linewidth]{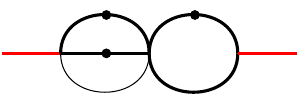}
		\caption*{$\M^{\text{NP1}}_{8}$}
	\end{minipage}
	\begin{minipage}{0.2\linewidth}
		\centering
		\includegraphics[width=1\linewidth]{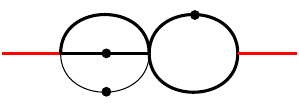}
		\caption*{$\M^{\text{NP1}}_{9}$}
	\end{minipage}
	\begin{minipage}{0.2\linewidth}
		\centering
		\includegraphics[width=1\linewidth]{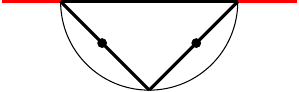}
		\caption*{$\M^{\text{NP1}}_{10}$}
	\end{minipage}
	\begin{minipage}{0.2\linewidth}
		\centering
		\includegraphics[width=1\linewidth]{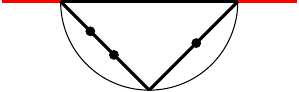}
		\caption*{$\M^{\text{NP1}}_{11}$}
	\end{minipage}
	\begin{minipage}{0.2\linewidth}
		\centering
		\includegraphics[width=1\linewidth]{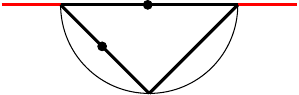}
		\caption*{$\M^{\text{NP1}}_{12}$}
	\end{minipage}
	\begin{minipage}{0.2\linewidth}
		\centering
		\includegraphics[width=1\linewidth]{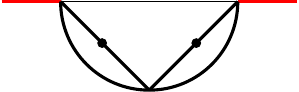}
		\caption*{$\M^{\text{NP1}}_{13}$}
	\end{minipage}
	\begin{minipage}{0.2\linewidth}
		\centering
		\includegraphics[width=1\linewidth]{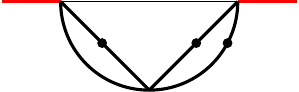}
		\caption*{$\M^{\text{NP1}}_{14}$}
	\end{minipage}
	\begin{minipage}{0.2\linewidth}
		\centering
		\includegraphics[width=1\linewidth]{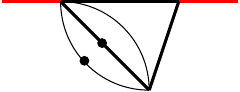}
		\caption*{$\M^{\text{NP1}}_{15}$}
	\end{minipage}
	\begin{minipage}{0.2\linewidth}
		\centering
		\includegraphics[width=1\linewidth]{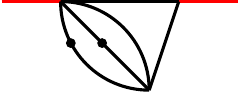}
		\caption*{$\M^{\text{NP1}}_{16}$}
	\end{minipage}
	\begin{minipage}{0.2\linewidth}
		\centering
		\includegraphics[width=1\linewidth]{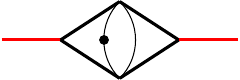}
		\caption*{$\M^{\text{NP1}}_{17}$}
	\end{minipage}
	\begin{minipage}{0.2\linewidth}
		\centering
		\includegraphics[width=1\linewidth]{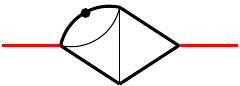}
		\caption*{$\M^{\text{NP1}}_{18}$}
	\end{minipage}
	\begin{minipage}{0.2\linewidth}
		\centering
		\includegraphics[width=1\linewidth]{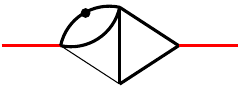}
		\caption*{$\M^{\text{NP1}}_{19}$}
	\end{minipage}
	\begin{minipage}{0.2\linewidth}
		\centering
		\includegraphics[width=1\linewidth]{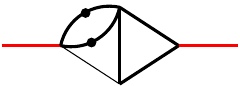}
		\caption*{$\M^{\text{NP1}}_{20}$}
	\end{minipage}
	\begin{minipage}{0.2\linewidth}
		\centering
		\includegraphics[width=1\linewidth]{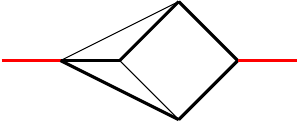}
		\caption*{$\M^{\text{NP1}}_{21}$}
	\end{minipage}
	\begin{minipage}{0.2\linewidth}
		\centering
		\includegraphics[width=1\linewidth]{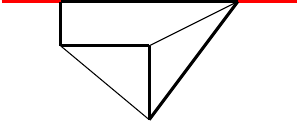}
		\caption*{$\M^{\text{NP1}}_{22}$}
	\end{minipage}
	\begin{minipage}{0.2\linewidth}
		\centering
		\includegraphics[width=1\linewidth]{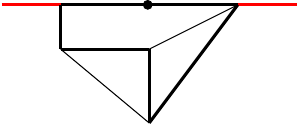}
		\caption*{$\M^{\text{NP1}}_{23}$}
	\end{minipage}
	\begin{minipage}{0.2\linewidth}
		\centering
		\includegraphics[width=1\linewidth]{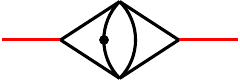}
		\caption*{$\M^{\text{NP1}}_{24}$}
	\end{minipage}
	\begin{minipage}{0.2\linewidth}
		\centering
		\includegraphics[width=1\linewidth]{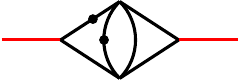}
		\caption*{$\M^{\text{NP1}}_{25}$}
	\end{minipage}
	\begin{minipage}{0.2\linewidth}
		\centering
		\includegraphics[width=1\linewidth]{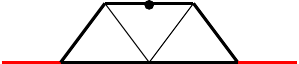}
		\caption*{$\M^{\text{NP1}}_{26}$}
	\end{minipage}
	\begin{minipage}{0.2\linewidth}
		\centering
		\includegraphics[width=1\linewidth]{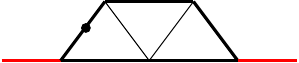}
		\caption*{$\M^{\text{NP1}}_{27}$}
	\end{minipage}
	\begin{minipage}{0.2\linewidth}
		\centering
		\includegraphics[width=1\linewidth]{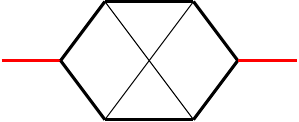}
		\caption*{$\M^{\text{NP1}}_{28}$, $\M^{\text{NP1}}_{29}$}
	\end{minipage}
\caption{Master integrals in the NP1 topology. The thick black and red lines stand for the massive bottom quark and the Higgs boson, respectively. One black dot indicates one additional power of the corresponding propagator. The numerators can be found in the text.}
\label{NP1_Topo}
\end{figure}

\begin{figure}[H]
	\centering
	\begin{minipage}{0.2\linewidth}
		\centering
		\includegraphics[width=1\linewidth]{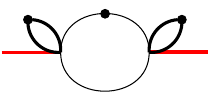}
		\caption*{$\M^{\text{P1}}_{1}$}
	\end{minipage}
	\begin{minipage}{0.2\linewidth}
		\centering
		\includegraphics[width=1\linewidth]{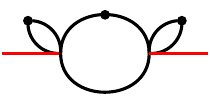}
		\caption*{$\M^{\text{P1}}_{2}$}
	\end{minipage}
	\begin{minipage}{0.2\linewidth}
		\centering
		\includegraphics[width=1\linewidth]{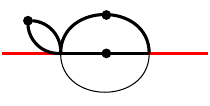}
		\caption*{$\M^{\text{P1}}_{3}$}
	\end{minipage}
	\centering
	\begin{minipage}{0.2\linewidth}
		\centering
		\includegraphics[width=1\linewidth]{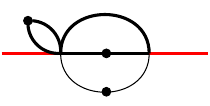}
		\caption*{$\M^{\text{P1}}_{4}$}
	\end{minipage}
	\begin{minipage}{0.2\linewidth}
		\centering
		\includegraphics[width=1\linewidth]{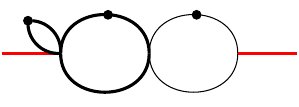}
		\caption*{$\M^{\text{P1}}_{5}$}
	\end{minipage}
	\begin{minipage}{0.2\linewidth}
		\centering
		\includegraphics[width=1\linewidth]{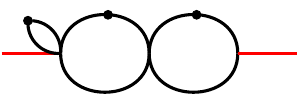}
		\caption*{$\M^{\text{P1}}_{6}$}
	\end{minipage}
	\begin{minipage}{0.2\linewidth}
		\centering
		\includegraphics[width=1\linewidth]{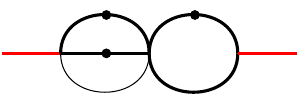}
		\caption*{$\M^{\text{P1}}_{7}$}
	\end{minipage}
	\begin{minipage}{0.2\linewidth}
		\centering
		\includegraphics[width=1\linewidth]{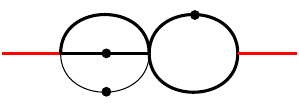}
		\caption*{$\M^{\text{P1}}_{8}$}
	\end{minipage}
	\begin{minipage}{0.2\linewidth}
		\centering
		\includegraphics[width=1\linewidth]{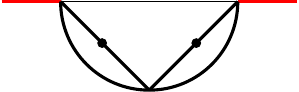}
		\caption*{$\M^{\text{P1}}_{9}$}
	\end{minipage}
	\begin{minipage}{0.2\linewidth}
		\centering
		\includegraphics[width=1\linewidth]{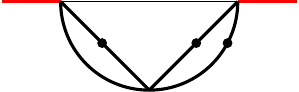}
		\caption*{$\M^{\text{P1}}_{10}$}
	\end{minipage}
	\begin{minipage}{0.2\linewidth}
		\centering
		\includegraphics[width=1\linewidth]{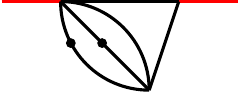}
		\caption*{$\M^{\text{P1}}_{11}$}
	\end{minipage}
	\begin{minipage}{0.25\linewidth}
		\centering
		\includegraphics[width=1\linewidth]{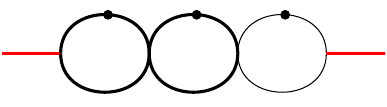}
		\caption*{$\M^{\text{P1}}_{12}$}
	\end{minipage}
	\begin{minipage}{0.2\linewidth}
		\centering
		\includegraphics[width=1\linewidth]{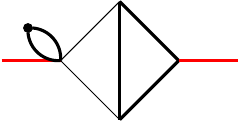}
		\caption*{$\M^{\text{P1}}_{13}$}
	\end{minipage}
	\begin{minipage}{0.2\linewidth}
		\centering
		\includegraphics[width=1\linewidth]{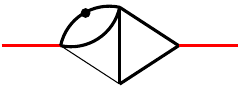}
		\caption*{$\M^{\text{P1}}_{14}$}
	\end{minipage}
	\begin{minipage}{0.2\linewidth}
		\centering
		\includegraphics[width=1\linewidth]{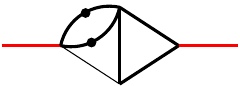}
		\caption*{$\M^{\text{P1}}_{15}$}
	\end{minipage}
	\begin{minipage}{0.2\linewidth}
		\centering
		\includegraphics[width=1\linewidth]{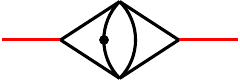}
		\caption*{$\M^{\text{P1}}_{16}$}
	\end{minipage}
	\begin{minipage}{0.2\linewidth}
		\centering
		\includegraphics[width=1\linewidth]{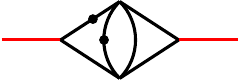}
		\caption*{$\M^{\text{P1}}_{17}$}
	\end{minipage}
	\begin{minipage}{0.2\linewidth}
		\centering
		\includegraphics[width=1\linewidth]{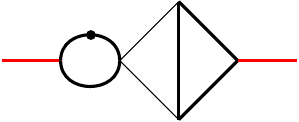}
		\caption*{$\M^{\text{P1}}_{18}$}
	\end{minipage}
	\begin{minipage}{0.2\linewidth}
		\centering
		\includegraphics[width=1\linewidth]{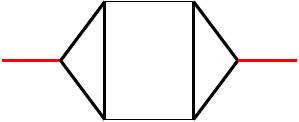}
		\caption*{$\M^{\text{P1}}_{19}$}
	\end{minipage}
\caption{Master integrals in the P1 topology. The thick black and red lines stand for the massive bottom quark and the Higgs boson, respectively. One black dot indicates one additional power of the corresponding propagator.}
\label{P1_Topo}
\end{figure}

The topology diagrams of the master integrals in the NP1 and P1
are displayed in figure \ref{NP1_Topo} and figure \ref{P1_Topo}.

\section{Asymptotic expansion of the master integrals}
\label{sec:asyMI}

In the limit of $z\rightarrow \infty\,(m_b \rightarrow 0 )$,
all the master integrals that are needed in $X_{2,b\bar{b}b\bar{b}}^{y_by_b}$  can be expanded  with the method in \cite{Lee:2020obg}. 
Here we present the asymptotic expansion results:
\begin{align}
&{F^{4b}_1}(z)|_{z\rightarrow \infty} =  12 \pi  \log ^2(z)-4 \pi ^3
\nonumber\\ &\quad
-\frac{1}{z}\bigg(144\pi\log^2(z)+144\pi\log(z)-48 \pi ^3\bigg)+\mathcal{O}(z^{-2})
,\nonumber\\
&{F^{4b}_2}(z)|_{z\rightarrow \infty} =  18 \pi  \log ^2(z)-24 \pi  \log (z)-6 \pi ^3
\nonumber\\ &\quad
-\frac{1}{z}\bigg(72\pi\log^2(z)+72\pi\log(z)-24 \pi ^3-144\pi\bigg)
+\mathcal{O}(z^{-2}),\nonumber\\
&{F^{4b}_3}(z)|_{z\rightarrow \infty} =  -6 \pi  \log (z)+12 \pi
\nonumber\\ &\quad
+\frac{1}{z}\bigg(36\pi\log^2(z)-36\pi\log(z)-12 \pi ^3+36\pi\bigg)
+\mathcal{O}(z^{-2}),\nonumber\\
&{F^{4b}_4}(z)|_{z\rightarrow \infty} =  12 \pi  \log ^2(z)-8 \pi ^3
\nonumber\\ &\quad
-\frac{1}{z}\bigg(72\pi\log^2(z)+48\pi\log(z)-24 \pi ^3+96\pi\bigg)
+\mathcal{O}(z^{-2}),\nonumber\\
&{F^{4b}_5}(z)|_{z\rightarrow \infty} =  4 \pi  \log ^3(z)-4 \pi ^3 \log (z)-96 \pi  \zeta (3)
\nonumber\\ &\quad
+\frac{1}{z}\bigg(144\pi\log^2(z)+432\pi\log(z)-48 \pi ^3+432\pi\bigg)
+\mathcal{O}(z^{-2}),\nonumber\\
&{F^{4b}_6}(z)|_{z\rightarrow \infty} =  -4 \pi  \log ^3(z)+8 \pi ^3 \log (z)-36 \pi  \zeta (3)-8 \pi ^3 \log (2)
\nonumber\\ &\quad
-\frac{1}{z}\bigg(24\pi\log^2(z)+96\pi\log(z)+8 \pi ^3\bigg)
+\mathcal{O}(z^{-2}),\nonumber\\
&{F^{4b}_7}(z)|_{z\rightarrow \infty} =  4 \pi  \log ^3(z)-8 \pi ^3 \log (z)+72 \pi  \zeta (3)
\nonumber\\ &\quad
+\frac{1}{z}\bigg(48\pi\log^2(z)+144\pi\log(z)-8 \pi ^3+48\pi\bigg)
+\mathcal{O}(z^{-2})
,\nonumber\\
&{F^{4b}_8}(z)|_{z\rightarrow \infty} =  -2 \pi  \log ^3(z)+6 \pi ^3 \log (z)-60 \pi  \zeta (3)-8 \pi ^3 \log (2)
\nonumber\\ &\quad
-\frac{1}{z}\bigg(12\pi\log^2(z)-4 \pi ^3-96\pi\bigg)
+\mathcal{O}(z^{-2}),\nonumber\\
&{F^{4b}_9}(z)|_{z\rightarrow \infty} =  -\frac{1}{2} \pi  \log ^4(z)+\pi ^3 \log ^2(z)+24 \pi  \zeta (3) \log (z)-\frac{19 \pi ^5}{10}
\nonumber\\ &\quad
+\frac{1}{z}\bigg(
4 \pi\log^3(z)+24\pi\log^2(z)-4\pi^3\log(z)+144\pi\log(z)-48\pi\zeta (3)+8 \pi ^3+240 \pi
\bigg)\nonumber\\ & \quad
+\mathcal{O}(z^{-2}),\nonumber\\
&{F^{4b}_{10}}(z)|_{z\rightarrow \infty}=  \pi  \log ^4(z)-4 \pi ^3 \log ^2(z)+72 \pi  \zeta (3) \log (z)+\frac{13 \pi ^5}{15}
\nonumber\\ &\quad
-\frac{1}{z}\bigg(48\pi\log^2(z)+240\pi\log(z)-8 \pi ^3+288\pi\bigg)
+\mathcal{O}(z^{-2}),\nonumber\\
&{F^{4b}_{12}}(z)|_{z\rightarrow \infty} =  \frac{3}{4} \pi  \log ^4(z)-\frac{7}{2} \pi ^3 \log ^2(z)+96 \pi  \zeta (3) \log (z)-\frac{43 \pi ^5}{60}
\nonumber\\ &\quad
-\frac{1}{z}\bigg(
6 \pi\log^3(z)+30\pi\log^2(z)-14\pi^3\log(z)+96\pi\log(z)+192\pi\zeta (3)-10 \pi ^3+36 \pi
\bigg)\nonumber\\
&\quad +\mathcal{O}(z^{-2}).
\end{align}

\bibliographystyle{JHEP}
\bibliography{Hbb}
\end{document}